\documentclass[aps,pra,reprint,amsmath,amssymb,twocolumn]{revtex4-1}

\usepackage{graphicx}% Include figure files
\usepackage{amssymb}
\usepackage{amsmath}
\usepackage{dcolumn}% Align table columns on decimal point
\usepackage{bm}% bold math
\usepackage{mathrsfs}% bold math
\usepackage{color}
\usepackage{hyperref}

\begin{document}

\title{The Unconventional Photon Blockade}

\author{H. Flayac}
\email{hugo.flayac@epfl.ch}
\author{V. Savona}
\email{vincenzo.savona@epfl.ch}
\affiliation{Institute of Physics iPHYS, \'{E}cole Polytechnique F\'{e}d\'{e}rale de Lausanne EPFL, CH-1015 Lausanne, Switzerland}

\begin{abstract}
We review the unconventional photon blockade mechanism. This quantum effect remarkably enables a strongly sub-Poissonian light statistics, even from a system characterized by a weak single photon nonlinearity. We revisit the past results, which can be interpreted in terms of quantum interferences or optimal squeezing, and show how recent developments on input-output field mixing can overcome the limitations of the original schemes towards passive and integrable single photon sources. We finally present some valuable alternative schemes for which the unconventional blockade can be directly adapted.
\end{abstract}
\maketitle

\section{Introduction}
Nonclassical statistics \cite{Strekalov2017} is a highly desirable feature offered by quantum photonic platforms \cite{OBrien2009} as it stands upstream from most single photon emission schemes \cite{Eisaman2011}. It is typically achieved in cavity QED by optimally coupling a strongly nonlinear degree of freedom, such as a two level emitter, to a harmonic cavity mode \cite{Birnbaum2005,Claudon2010,Gazzano2013,Madsen2014,Dory2017,Hamsen2017}. In this direction, conventional schemes rely on the so-called ``photon blockade'' (PB) mechanism \cite{Tian1992,Leonski1994,Imamoglu1997} by analogy to the famous Coulomb blockade effect \cite{Grabert2013}: The auxiliary element or the strongly nonlinear medium induces an sizable anharmonicity in the excitation ladder which shifts the $n>2$ photon states off-resonance. As a consequence, the cavity can only host one photon at a time, behaving as a ``photon turnstile'' \cite{Michler2000}. This feature is associated with a sub-Poissonian statistics and a photon antibunching \cite{Paul1982} resulting from a non-Gaussian state. The efficiency of the PB mechanism however requires the single photon nonlinearity $U$ to be at least larger than the mode linewidth (losses) $\kappa$ to fully suppress the unwanted transitions. While systems relying on the PB are steadily improving and currently report close to optimal figures of merit \cite{He2013,Gschrey2015,Somaschi2016}, they still pose major technological challenges in term of integrability and scalability \cite{Michler2017}.

Weakly nonlinear systems, characterized by $U \ll \kappa$, are far more natural and appear in many areas of photonics but also of condensed matter physics. Weak nonlinearities typically stem from the medium itself \cite{Ferretti2013} or a weakly coupled nonlinear element \cite{Snijders2016}. Contrary to what is commonly believed, small nonlinear energy shifts are actually a sufficient ingredient to build up sizable quantum correlations even under weak driving \cite{Foster2000}. The key requirement is to couple at least two degrees of freedom in order to assist quantum interferences between excitation pathways \cite{Carmichael1991,Bamba2011,Majumdar2012,Radulaski2017,Kamide2017}. In that framework, a strongly sub-Poissonian statistics can be achieved by means of a pair of driven dissipative resonators with an arbitrarily small single photon nonlinearity. This effect is referred to as the ``unconventional photon blockade'' (UPB) \cite{Liew2010,Bamba2011,Flayac2013,Xu2014} and was originally thought for Kerr resonators, namely a Bose-Hubbard dimer, but it can be easily transposed to various configurations. As opposed to its conventional counterpart, the UPB relies on close to Gaussian states in the regime of weak nonlinearity. The effect can therefore be interpreted in terms of optimally squeezed states \cite{Lemonde2014}.

The UPB is a strongly resonant effect which, unlike parametric down-conversion \cite{Mosley2008} or four wave mixing \cite{Davancco2012} involved in heralded schemes \cite{Silverstone2016}, requires a minimum input intensity to operate \cite{Flayac2015}. The drawback is to work with intracavity fields below unity occupation and to accept a probabilistic single photon emission. Yet, the UPB is a very promising mechanism for integrable and scalable single photon sources since it doesn't require any quantum emitter to operate. It could be suitably applied to small footprint optimized Silicon photonic crystal cavities where the $\chi^{(3)}$ response naturally offers a weakly nonlinear Kerr medium \cite{Ferretti2013,Dharanipathy2014,Flayac2015}. Beyond single photons applications, the UPB can be used as a tool to reveal nonclassical features. For example, the thriving field of semiconductor microcavities \cite{Laussy2017} is now actively seeking for genuine quantum correlations \cite{Sanvitto2016}. Given the small single particle nonlinearity offered by excitons, a nonclassical light statistics can hardly be observed from the exciton-polariton field without relying on the UPB \cite{Flayac2017}.

Yet, there are two longstanding obstacles that have prevented the experimental realization of the UPB in its original form: (i) It requires a fine tuning of the intrinsic system parameters as the optimal sub-Poissonian statistics is obtained for a proper interrelations between the cavity coupling, the nonlinearity and the laser detuning. (ii) A weak nonlinearity imposes a large coupling between the two cavities which, in turn, results in fast oscillations of the second order correlation function on a time scale smaller than the cavity lifetime \cite{Bamba2011,Flayac2016}. As a result, the sub-Poissonian window of the UPB is difficult to extract within the temporal resolution of state-of-the-art detectors, and pulsed operation even turns out to produce super-Poissonian light.

These issues actually arose due to the initial formulation of the model where the authors imposed to drive only one of the cavity modes \cite{Liew2010,Bamba2011}. Indeed, by allowing a mutual driving of the modes and/or a mixing of their output, we will show that the parameter constraints are naturally absorbed in the relative phase and amplitude of the coherent sources that can be tuned at will in experiments. As a consequence, one can achieve a strongly sub-Poissonian statistics associated with a well behaved second order correlation function for a wide range of cavity parameters.

The manuscript is organized as follows: In Sec.\ref{Sec:Formalism} we introduce the general formalism and the mathematical tools for the quantum description of the system. We review the original proposals in Sec.\ref{Sec:OriginalUPB}. In Sec.\ref{Sec:Squeezing}, we present the interpretation of the UPB in terms of optimal squeezing. In Sec.\ref{Sec:Developments}, we discuss the latest developments and present new schemes that may lead to an experimental evidence of the UPB. Finally in the conclusions of Sec.\ref{Sec:Conclusions}, we discuss the outlooks in terms of applications and variations of the UPB.

\begin{figure}[ht]
\includegraphics[width=0.45\textwidth,clip]{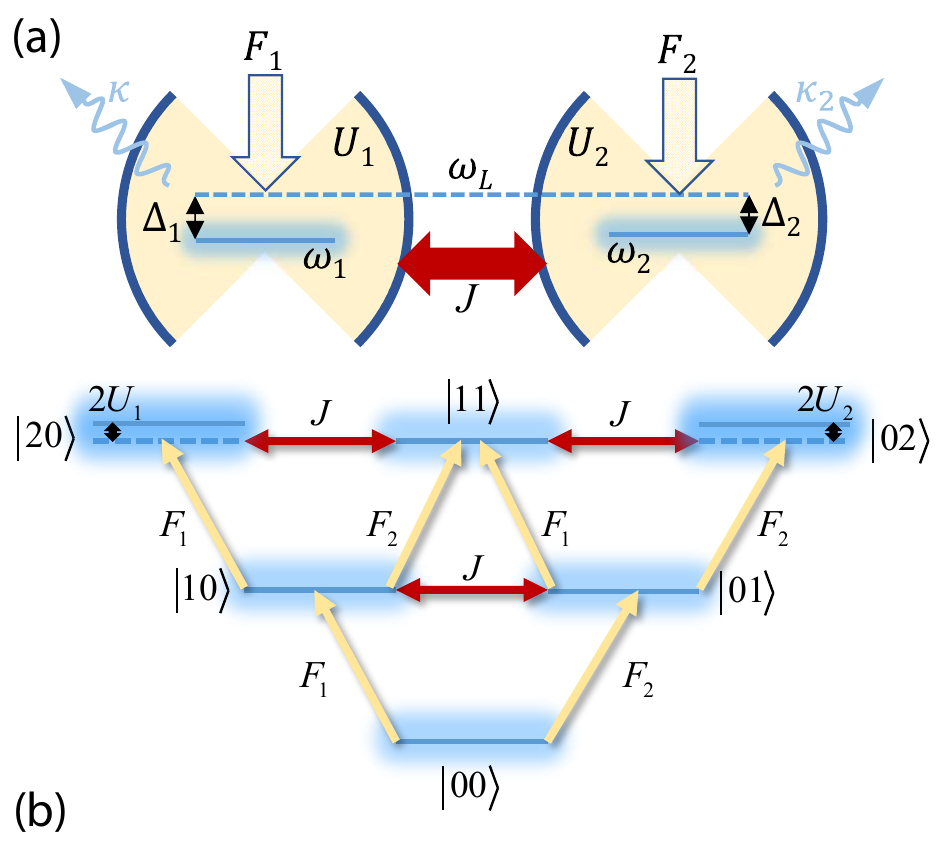}\\
\caption{(Color online) (a) Scheme of the system: Coupled Kerr cavities in a driven-dissipative environment. The modes of linewidth $\kappa_{1,2}$ allow for photons hopping with an amplitudes $J$ and are excited by mutually coherent classical sources of complex amplitudes $F_{1,2}$. (b) Energy levels in the two-photon manifold and the corresponding excitation paths.}
\label{Fig1}
\end{figure}

\section{General Formalism}\label{Sec:Formalism}
We shall consider here the general model of two coupled single-mode cavities with resonant frequencies $\omega_j$ ($j=1,\,2$), both containing a Kerr medium, which are driven by classical sources as sketched in Fig.\ref{Fig1}(a). In the frame rotating at the laser frequency $\omega_L$, the system Hamiltonian reads
\begin{eqnarray}\label{H}
\nonumber \hat {\cal H} &=& \sum\limits_{j = 1,2} {[ {{\Delta _j}\hat a_j^\dag {{\hat a}_j} + {U_j}\hat a_j^{\dag 2}\hat a_j^2 + F_j^*{{\hat a}_j} + {F_j}\hat a_j^\dag } ]}  \hfill \\
   &+& J \left( {\hat a_1^\dag {{\hat a}_2} + \hat a_2^\dag {{\hat a}_1}} \right)
\end{eqnarray}
Here $\Delta_{j}=\omega_{j}-\omega_L$ are the cavity detunings, $F_{j}$ the complex amplitudes of the driving fields, $U_{j}$ are the Kerr nonlinearity magnitudes, and $J$ is the hopping amplitude between the two cavities.

\subsection{Weak driving limit}\label{Sec:WeakDriving}
Before introducing the full treatment of the open quantum system, we present here a simplified description in terms of pure states and in the limit of weak driving fields. In this limiting case, we derive analytical expressions for the system observables and find optimal conditions for the system parameters that maximize the sub-Poissonian character of the cavity field \cite{Bamba2011,Flayac2013,Flayac2016}. We shall express the quantum state as an expansion on the basis of occupation number eigenstates. In the limit where $F_{1,2}\rightarrow0$, it is then possible to retain only terms in this expansion, whose coefficients depend to leading order in the driving field amplitudes. From the Schr\"{o}dinger equation, it can be inferred that the coefficient $c_{nm}$ depends exactly as ${\cal{O}}( {F_1^n F_2^m})$ to leading order. Hence, in the weak driving limit, the time-dependent state is well approximated in the 2-photon manifold as
\begin{equation}\label{psit}
\left| {\psi \left( t \right)} \right\rangle  \simeq \sum\limits_{n+m\leq2} {{c_{nm}}\left( t \right)} \left| {nm} \right\rangle
\end{equation}
Here $\left| nm \right\rangle$ denotes a state with $n$ photons in the first cavity and $m$ photons in the second one. In the most general case, the state \eqref{psit} should evolve according to a stochastic Schr\"{o}dinger equation, including random quantum jumps \cite{Dum1992,Molmer1993,Carmichael2007}. For vanishing occupation however, in the presence of losses at rates $\kappa_j$, the time evolution of the state \eqref{psit} is well approximated by its ``jumpless'' description as quantum jumps become extremely rare. Hence, the equations governing the time-dependence of the coefficients are found from the solution of the Schr\"{o}dinger equation $\tilde {\cal H}\left|\psi\right\rangle=i\hbar{\partial _t}\left| \psi  \right\rangle$ written for the non-Hermitian Hamiltonian
\begin{equation}\label{HNH}
\tilde {\cal{H}} = \hat{\cal{H}} - i\sum\limits_{j = 1,2} {\frac{\kappa_j}{2}\hat a_j^\dag {{\hat a}_j}}
\end{equation}
The equations for the $c_{nm}(t)$ are obtained by projection on the $\left| {nm} \right\rangle$ states and in particular
\begin{eqnarray}
\label{c00}
  i{{\dot c}_{00}} &=& \underline{F_1^* c_{10}} + \underline{F_2^* c_{01} \hfill} \\
\label{c10}
  i{{\dot c}_{10}} &=& {F_1}{c_{00}} + {\tilde\Delta _1}{c_{10}} + J{c_{01}} + \underline{F_2^*{c_{11}}} + \underline{F_1^*\sqrt 2 {c_{20}}} \hfill \\
\label{c01}
  i{{\dot c}_{01}} &=& {F_2}{c_{00}} + {\tilde\Delta _2}{c_{01}} + J{c_{10}} + \underline{F_1^*{c_{11}}} + \underline{F_2^*\sqrt 2 {c_{02}}} \hfill \\
\label{c20}
  i{{\dot c}_{20}} &=& {F_1}\sqrt 2 {c_{10}} + J\sqrt 2 {c_{11}} + 2( {{\tilde\Delta _1} + {U_1}} ){c_{20}} \hfill \\
\label{c02}
  i{{\dot c}_{02}} &=& {F_2}\sqrt 2 {c_{01}} + J\sqrt 2 {c_{11}} + 2( {{\tilde\Delta _2} + {U_2}} ){c_{02}} \hfill \\
  \nonumber  i{{\dot c}_{11}} &=& {F_2}{c_{10}} + {F_1}{c_{01}} + ( {{\tilde\Delta _1} + {\tilde\Delta _2}} ){c_{11}} + J\sqrt 2 {{c_{20}} + J\sqrt 2{c_{02}}} \hfill \\
\label{c11}
\end{eqnarray}
with the definition $\tilde{\Delta}_j\equiv\Delta_j-i\kappa_j/2$. The underlined terms in Eqs.(\ref{c00}-\ref{c01}) are of sub-leading order in the driving field amplitudes and can be neglected. The figure \ref{Fig1}(b), showing the energy levels and the links between the states imposed by Eq.\eqref{HNH}, directly illustrates the set (\ref{c00}-\ref{c11}). Under continuous wave driving, the equations are solved for the steady state $\left| {\psi_{\rm ss}} \right\rangle$ where $\dot{c}_{nm}(t)=0$. Given that $c_{00} \gg c_{10},c_{10} \gg c_{20},c_{02},c_{11}$, we impose the condition $c_{00}=1$ which compensates for the small decay of the norm induced by Eq.\eqref{HNH} and allows a simple closing of the equations. The Eqs.(\ref{c10},\ref{c11}) are solved recurrently and allow obtaining compact expressions for the $c_{nm}$. Then, the average occupations and equal-time second order correlations approximate to
\begin{eqnarray}
\label{n1}
{n_1} = \langle {\hat a_1^\dag {{\hat a}_1}} \rangle &=& \left| c_{10} \right|^2 + \left|c_{11}\right|^2 + 2\left|c_{20}\right|^2 \simeq \left| c_{10} \right|^2 \hfill \\
\label{n2}
{n_2} = \langle {\hat a_2^\dag {{\hat a}_2}} \rangle &=& \left| c_{01} \right|^2 + \left|c_{11}\right|^2 + 2\left|c_{02}\right|^2 \simeq \left| c_{01} \right|^2 \hfill \\
\label{g201}
g_1^{(2)}(0) &=& \frac{{\langle \hat a_1^\dag \hat a_1^\dag {{\hat a}_1}{{\hat a}_1}\rangle }}{{n_1^2}} \simeq 2\frac{{{{\left| {{c_{20}}} \right|}^2}}}{{{{\left| {{c_{10}}} \right|}^4}}} \\
\label{g202}
g_2^{(2)}(0) &=& \frac{{\langle \hat a_2^\dag \hat a_2^\dag {{\hat a}_2}{{\hat a}_2}\rangle }}{{n_2^2}} \simeq 2\frac{{{{\left| {{c_{02}}} \right|}^2}}}{{{{\left| {{c_{01}}} \right|}^4}}}
\end{eqnarray}
In the general case of a $n+m$-photon manifold, the coefficients $c_{nm}$ are determined by the recurrence relation
\begin{eqnarray}
\nonumber i{{\dot c}_{nm}} &=& {{\tilde \Delta }_{nm}}{c_{nm}} + {F_1}\sqrt n {c_{n - 1m}} + {F_2}\sqrt m {c_{nm - 1}} \hfill \\
                           &+& F_1^*\sqrt {n + 1} {c_{n + 1m}} + F_2^*\sqrt {m + 1} {c_{nm + 1}} \hfill \\
\nonumber                  &+& J\sqrt {n\left( {m + 1} \right)} {c_{n - 1m + 1}} + J\sqrt {m\left( {n + 1} \right)} {c_{n + 1m - 1}}
\end{eqnarray}
where ${{\tilde \Delta }_{nm}} \equiv n{{\tilde \Delta }_1} + m{{\tilde \Delta }_2} + n\left( {n - 1} \right){U_1} + m\left( {m - 1} \right){U_2}$. The second order correlations read
\begin{eqnarray}
g_1^{\left( 2 \right)}(0) &=& \frac{1}{{n_1^2}}\sum\limits_{n,m} {n\left( {n - 1} \right)|{c_{nm}}{|^2}} \hfill \\
g_2^{\left( 2 \right)}(0) &=& \frac{1}{{n_2^2}}\sum\limits_{n,m} {m\left( {m - 1} \right)|{c_{nm}}{|^2}}
\end{eqnarray}
given the mean occupancies $n_1=\sum\nolimits_{n,m} {n|{c_{nm}}{|^2}}$ and $n_2=\sum\nolimits_{n,m} {m|{c_{nm}}{|^2}}$.

\subsection{Numerical Treatment}
To correctly account for the driven-dissipative character of the system, we introduce the quantum master equation for the system density matrix
\begin{equation}\label{rhot}
  i\frac{{\partial \hat \rho }}{{\partial t}} =  \left[ {\hat {\cal{H}}},\hat \rho\right] - i{\sum\limits_{j = 1,2} \frac{{{\kappa_j}}}{2}\hat {\cal{D}}\left[ {{{\hat a}_j}} \right]\hat \rho} \,.
\end{equation}
Here, $\hat {\cal{D}}\left[ {{{\hat a}_j}} \right]\hat \rho=\{\hat a_j^\dag {{\hat a_j}},\hat \rho\} - 2{{\hat a_j}}\hat \rho \hat a_j^\dag$ are Lindblad terms accounting for losses to the environment. The expectation values are computed as $\langle\hat o\rangle={\rm{Tr}}(\hat o\hat\rho)$. In what follows, we will derive numerical solutions of Eq.(\ref{rhot}) in a truncated Hilbert space where only states $|nm\rangle$ with $n+m\le N_{\rm max}$ are retained, and the convergence of the results vs $N_{\rm max}$ is carefully checked. With this approach, cases with moderate driving field amplitude can be accurately modeled.

For still larger driving fields, the relevant occupation numbers are such that the above approach becomes numerically cumbersome. In this limit however, we expect the field in the two cavity modes to be well described by small quantum fluctuations occurring on classical field amplitudes. It is then possible to expand the photon operators as $\hat a_j = \alpha_j\hat {\mathbb{I}} + \delta \hat a_j$, where $\alpha_j=\langle \hat a_j \rangle$ is the coherent mean field component and $\delta \hat a_j$ are the fluctuation (noise) operators fulfilling $\langle \delta\hat a_j \rangle\approx0$ \cite{Flayac2016,Flayac2017}. The classical field dynamics follows
\begin{eqnarray}
\label{alpha1t}  i{\dot \alpha}_1 &=& [\tilde \Delta_1 + U_1\left| \alpha_1 \right|^2]{\alpha _1} + J\alpha_2 + F_1\\
\label{alpha2t}  i{\dot \alpha}_2 &=& [\tilde \Delta_2 + U_2\left| \alpha_2 \right|^2]{\alpha _2} + J\alpha_1 + F_2
\end{eqnarray}
and the fluctuations are governed by the master equation
\begin{equation}\label{rhotf}
  i\hbar\frac{{\partial \hat \rho_f }}{{\partial t}} =  \left[ {\hat {\cal{H}}_f,\hat \rho_f } \right] - i{\sum\limits_{j = 1,2} \frac{{{\kappa_j}}}{2}\hat {\cal{D}}\left[ {{{\delta\hat a}_j}} \right]\hat \rho_f}
\end{equation}
The corresponding semiclassical Hamiltonian reads
\begin{eqnarray}\label{Hf}
{{\hat {\cal H}}_f} &=& \sum\limits_{j = 1,2} {\left[ {\Delta_j\hat a_j^\dag {{\hat a}_j} + {U_j}( {\alpha _j^{2*}\hat a_j^2 + \alpha _j^2\hat a_j^{\dag 2}} )} \right]}\\
\nonumber  &+& \sum\limits_{j = 1,2} {{U_j}\left[ {\hat a_j^\dag \hat a_j^\dag {{\hat a}_j}{{\hat a}_j} + 2\alpha _j^*\hat a_j^\dag {{\hat a}_j}{{\hat a}_j} + 2{\alpha _j}\hat a_j^\dag \hat a_j^\dag {{\hat a}_j}} \right]} \hfill \\
\nonumber &+& J\left(\hat a_1^\dag {{\hat a}_2} + \hat a_2^\dag {{\hat a}_1}\right)
\end{eqnarray}
We have voluntarily omitted the $\delta$ notation in Eq.\eqref{Hf} for the sake of compactness. This approach, where nonlinear fluctuation terms of all orders are kept, provides an exact description of the quantum dynamics as long as $\langle \delta\hat a_j \rangle\ll\alpha_j$. The expectation values are then computed as $\langle\delta\hat o + \langle \hat o \rangle \hat {\mathbb{I}}\rangle={\rm{Tr}}[(\delta\hat o + \langle \hat o \rangle \hat {\mathbb{I}}\hat\rho_f]$.

\section{Original proposal}\label{Sec:OriginalUPB}
\subsection{Photon statistics under continuous wave driving}\label{Sec:CWDrive}
In the works by Liew and Savona \cite{Liew2010} and Bamba et al. \cite{Bamba2011}, the analysis was restricted to the case where only one of the quantum modes is driven, namely $F_2=0$. Identical cavities where $\Delta_1=\Delta_2=\Delta$, $U_1=U_2=U$ and $\kappa_1=\kappa_2=\kappa$ were also considered for simplicity. Under these simplifying assumptions, the coefficients of Eq.\eqref{psit} are found to be
\begin{eqnarray}
{c_{10}} &=& {F_1}\frac{{\tilde \Delta }}{{{J^2} - {{\tilde \Delta }^2}}} \hfill \\
{c_{01}} &=&  - {F_1}\frac{J}{{{J^2} - {{\tilde \Delta }^2}}} \hfill \\
\label{c20UPB}
\nonumber {c_{20}} &=& F_1^2\frac{{{J^2}U + 2{{\tilde \Delta }^2}\left[ {U + \tilde \Delta } \right]}}{{2\sqrt 2 \left[ {U + \tilde \Delta } \right]\left[ {{{\tilde \Delta }^2} - {J^2}} \right]\left[ {\tilde \Delta ( {U + \tilde \Delta } ) - {J^2}} \right] }} \hfill \\ \\
\nonumber {c_{02}} &=& F_1^2{J^2}\frac{{U + 2\tilde \Delta }}{{2\sqrt 2 \left[ {U + \tilde \Delta } \right]\left[ {{{\tilde \Delta }^2} - {J^2}} \right]\left[ {\tilde \Delta ( {U + \tilde \Delta } ) - {J^2}} \right]}} \hfill \\ \\
{c_{11}} &=& F_1^2J\frac{{U + 2\tilde \Delta }}{{2\left[ {{J^2} - {{\tilde \Delta }^2}} \right]\left[ {\tilde \Delta ( {U + \tilde \Delta } ) - {J^2}} \right]}}
\end{eqnarray}
The sub-Poissonian character of the cavity 1 statistics can then be optimized by solving for $c_{20}=0$ as prescribed by Eq.\eqref{g201} provided that $c_{10}\neq0$.

\begin{figure}[ht]
\includegraphics[width=0.49\textwidth,clip]{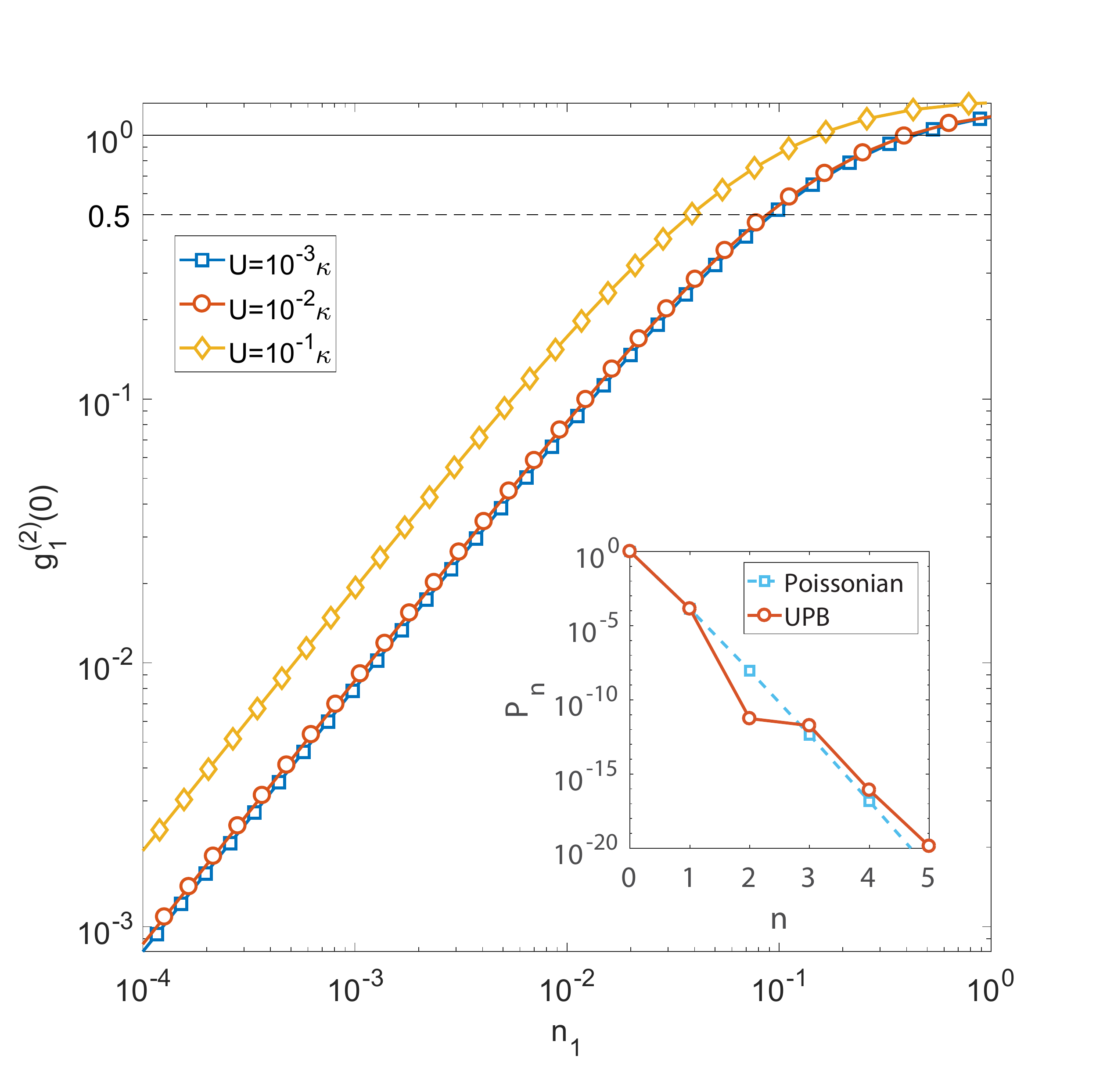}\\
\caption{(Color online) Equal time second order correlations of the driven cavity $g^{(2)}_1(0)$ versus its mean occupancy $n_1$ under the optimal conditions (\ref{Deltaopt},\ref{Uopt}) for several values of $U$ (see legend). The inset shows the corresponding probability distribution compared to a Poissonian statistics for $U=10^{-2}\kappa$ and $n_1\simeq10^{-3}$.}
\label{g21_vs_n1}
\end{figure}

In Ref.\cite{Bamba2011}, the following relations for optimal UPB were derived
\begin{eqnarray}
\label{Deltaopt}
  {\Delta _{|\rm{opt}}} &=&  \pm \frac{1}{2}\sqrt {\sqrt {9{J^4} + 8{\kappa ^2}{J^2}}  - {\kappa ^2} - 3{J^2}} \hfill \\
\label{Uopt}
  {U_{|\rm{opt}}} &=& \frac{{{\Delta _{|\rm{opt}}}\left( 4\Delta _{|\rm{opt}}^2 + {5{\kappa ^2}} \right)}}{{2\left( {2{J^2} - {\kappa ^2}} \right)}} \mathop  \simeq \limits^{U \ll \kappa } \frac{2}{{3\sqrt 3 }}\frac{{{\kappa ^2}}}{{{J^2}}}
\end{eqnarray}
resulting in a value of $g^{(2)}_1(0)=0$ associated with a perfectly destructive quantum interference between the direct excitation path $\left| {00} \right\rangle  \to \left| {10} \right\rangle  \to \left| {20} \right\rangle$ and the longest paths involving the second cavity $\left| {00} \right\rangle  \to \left| {10} \right\rangle  \to \left| {01} \right\rangle  \to \left( {\left| {11} \right\rangle  \leftrightarrow \left| {02} \right\rangle } \right) \to \left| {20} \right\rangle$ as it can be deduced from Fig.\ref{Fig1}(b) and Eq.\eqref{c20}. Indeed, the destructive interference occurs when the $\left| {10} \right\rangle$ and $\left| {11} \right\rangle$ contributions to $\left| {20} \right\rangle$ exactly cancel each other. Strictly speaking, this holds true only in the framework of the truncation of Eq.\eqref{psit}. Accounting for the $n+m>2$ states and/or allowing for state mixedness results in a small finite value for the equal-time second-order correlation function. To illustrate this point, we show in Fig.\ref{g21_vs_n1} the dependence of $g^{(2)}_1(0)$ on the cavity mean occupancy $n_1$ obtained by solving Eq.\eqref{rhotf} with optimal UPB conditions \eqref{Deltaopt} and \eqref{Uopt}, for different values of the optical nonlinearity. The result shows a linear increase of $g^{(2)}_1(0)$ at low occupancy $n_1$, which becomes nonlinear when approaching unit occupancy. It differs significantly from the conventional Kerr blockade case where the $g^{(2)}(0)\simeq|\tilde \Delta {|^4}/|\tilde \Delta (2U + \tilde \Delta ){|^2}$ function is instead constant for $n\ll1$. We also see from Fig.\ref{g21_vs_n1} that the smaller the nonlinearity, the smaller the $g^{(2)}_1(0)$ for a given occupancy. Remarkably, the single photon regime usually characterized by $g^{(2)}_1(0)<0.5$ is guaranteed up to $n_1\simeq0.1$ for $U=10^{-2}-10^{-3}\kappa$. We note that the analytical criterion provides the smallest $g^{(2)}_1(0)$ value possible for every occupancy $n_1\leq1$ as we have checked using a global minimization routine over $U$ and $J$. It indicates that the suppression of the two-photon probability is the best strategy to optimize the sub-Poissonian statistics for the UPB. The inset of Fig.\ref{g21_vs_n1} shows the photon probability distribution $P_n$ for $U=10^{-2}\kappa$ and $n_1\simeq10^{-3}$ compared to a Poissonian distribution with the same average photon number. We see the clear suppression by several orders of magnitudes of the 2-photon probability induced by the UPB while the $n>2$ probabilities are slightly enhanced. An interesting quantity to compute, in view of single photon applications, is the probability of emitting more than one photon $P_{n>1}$. In the present case, the Poissonian statistics produces 1000 times more multiphoton events on average for the same value of $P_1$.

\begin{figure}[ht]
\includegraphics[width=0.49\textwidth,clip]{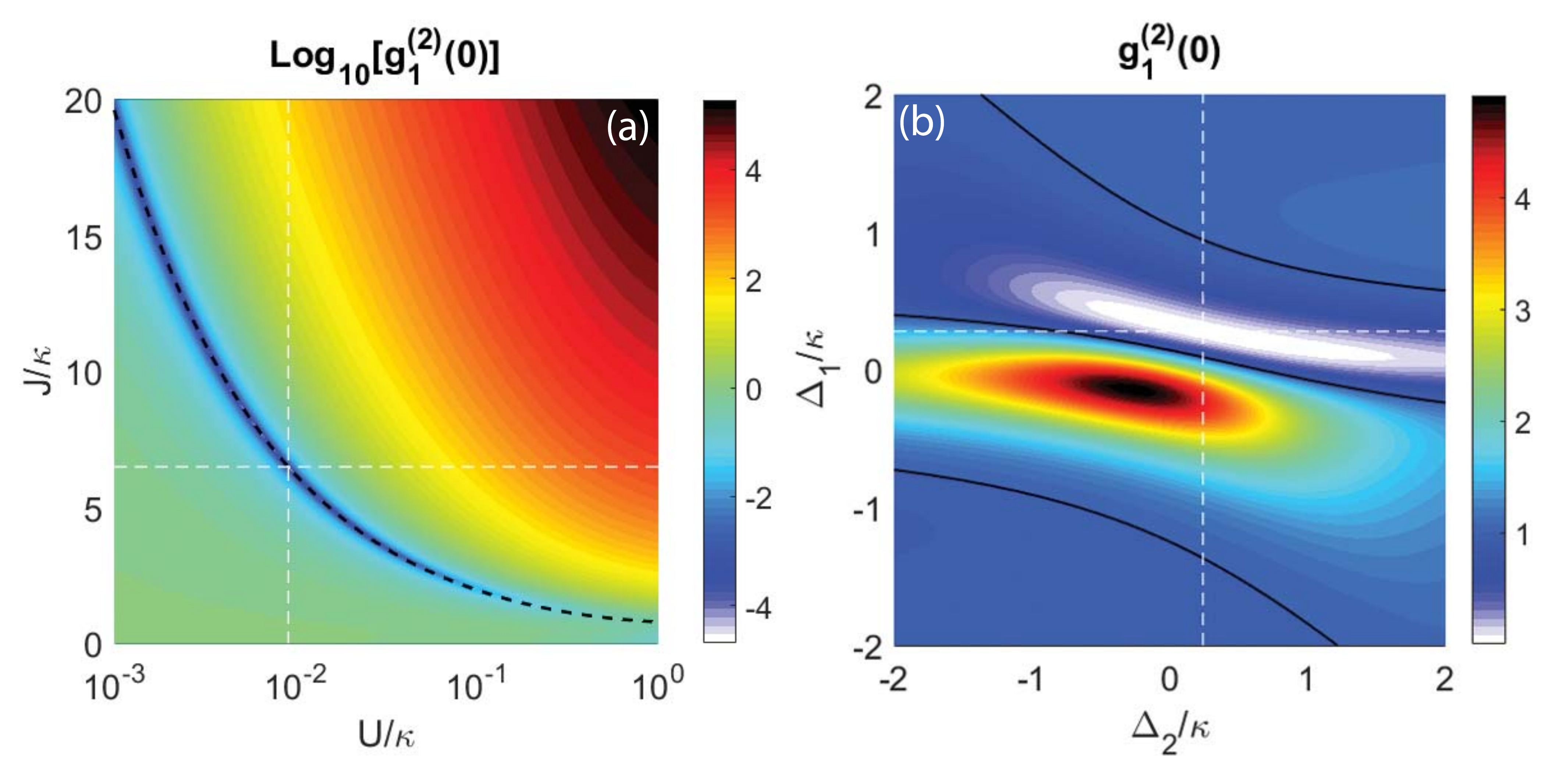}\\
\caption{(Color online) Equal time second order correlations of the driven cavity $g^{(2)}_1(0)$ versus $U$ and J and (b) versus $\Delta_1$ and $\Delta_2$. The dashed-white lines spot the minima of the $g^{(2)}_1(0)$ functions. The dashed black line mark the optimal link between $J$ and $U$.}
\label{g2_maps}
\end{figure}

In figure \ref{g2_maps}(a) we show a full $g^{(2)}_1(0)$ map versus $U$ and $J$ for $F_1=0.1\kappa$. The dashed black line highlights the optimal relation between $J$ and $U$ and the white lines mark the global minimum.
We additionally show in Fig.\ref{g2_maps}(b) the impact of variable detuning around the optimal condition $\Delta_{1}=\Delta_{2}=\Delta_{|\rm opt}$. The map reveals the strongly resonant character of the unconventional photon blockade induced by the underlying quantum interference mechanism. Note that by adjusting the detuning, one can also prepare the system for a super-Poissonian statistics [see dark-red areas] associated with a suppression of the $c_{10}$ coefficient.

\begin{figure}[ht!]
\includegraphics[width=0.49\textwidth,clip]{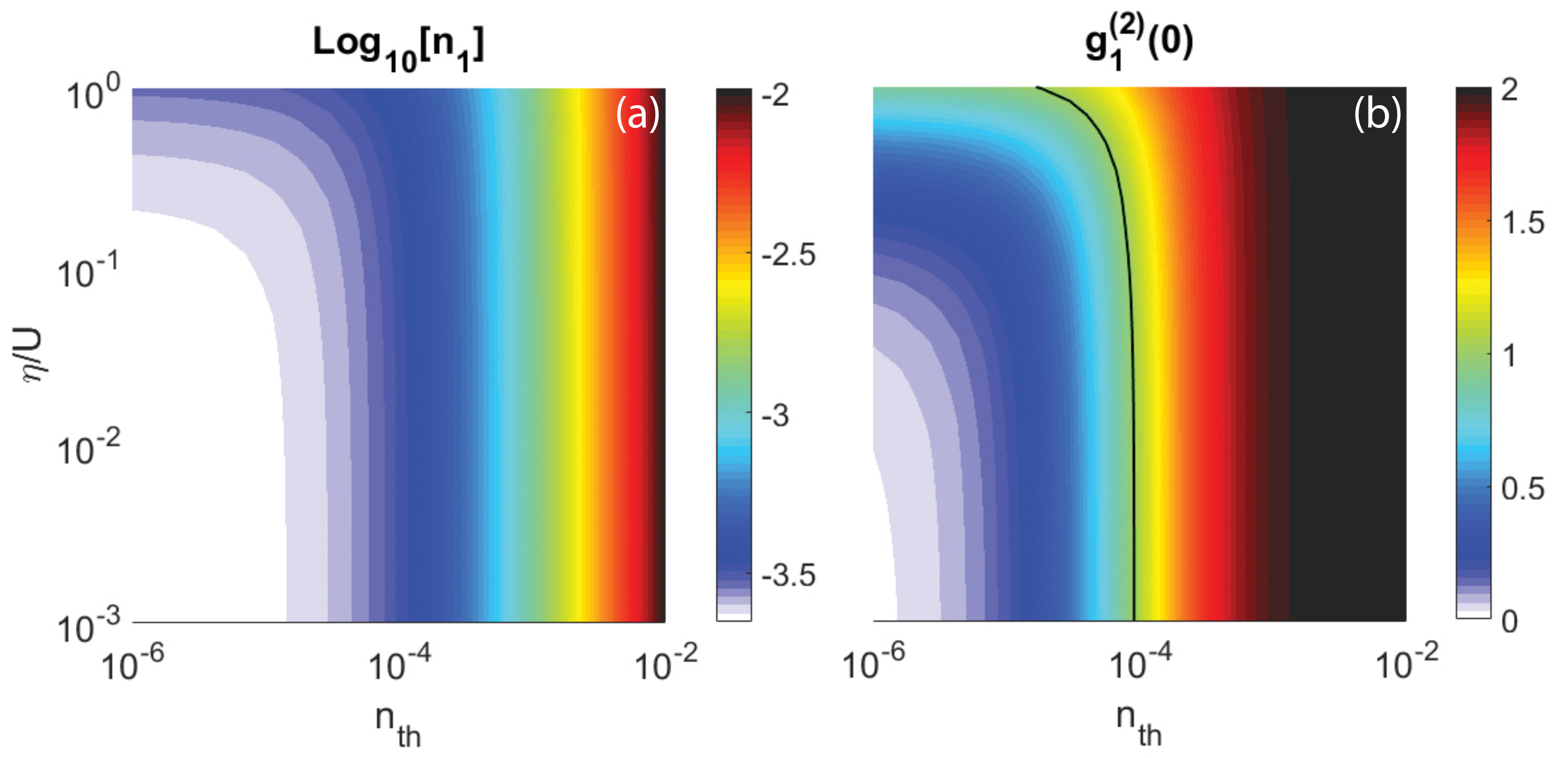}\\
\caption{(Color online) (a) Cavity 1 mean occupancy (log scale) and (b) equal time second order correlation function versus the mean thermal occupation $\bar n_{\rm th}$ and the pure dephasing rate $\eta$. The optimal condition is set for $U=10^{-2}\kappa$ and we fix $F_1=\kappa$.}
\label{g21_vs_nth_and_pd}
\end{figure}

Any realistic implementation of the mechanism will suffer from some form of decoherence mechanism which affects the quantum interference. Another possible detrimental mechanism may arise in systems with frequencies in the microwave range, such as superconducting circuits \cite{Eichler2014,Gu2017}, where finite temperature may induce a non-negligible thermal occupancy $\bar n_{\rm th}$. Thermal photons set a lower bound on the coherent contribution to the occupancy needed to overcome the thermal statistics leading to $g^{(2)}_1(0)=2$. The thermal contribution requires one to consider a gain of excitations from the reservoir. The Linblad terms of Eq.\eqref{rhot} are therefore rewritten as $- i\left( {{{\bar n}_{\rm th}} + 1} \right)\sum\nolimits_{j} {{{{\kappa _j}}}/{2}} \hat {\cal D}[ {{{\hat a}_j}} ]\hat \rho  - i{\bar n_{\rm th}}\sum\nolimits_{j} {{{{\kappa _j}}}/{2}} \hat {\cal D}[ {\hat a_j^\dag } ]\hat \rho$ where $\bar n_{\rm th}$ follows a Bose distribution. Pure dephasing is accounted for through the additional term $- i\eta/4 \sum\nolimits_j {{\cal D}[\hat a_j^\dag {{\hat a}_j}]} \hat \rho$. We show in Fig.\ref{g21_vs_nth_and_pd} a map of the emission statistics versus $\bar n_{\rm th}$ and the pure dephasing rate $\eta$. A smooth transition of the statistics occurs from sub-Poissonian (i) to thermal with increasing $\bar n_{\rm th}$ and (ii) to Poissonian when $\eta$ approaches $U$.

\begin{figure}[ht]
\includegraphics[width=0.49\textwidth,clip]{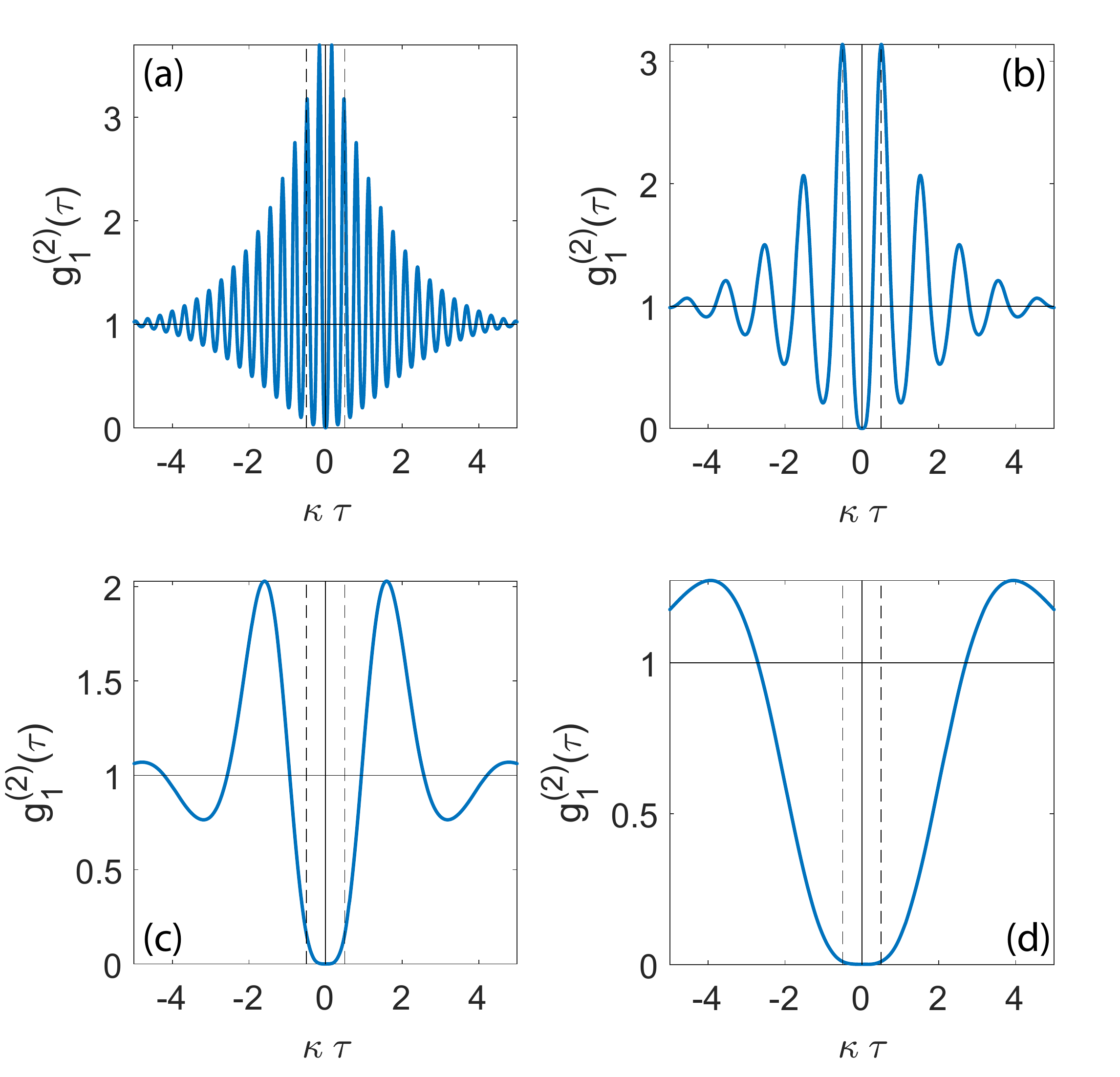}\\
\caption{(Color online) Delayed second order correlations of the driven cavity $g^{(2)}_1(\tau)$ under the optimal conditions (\ref{Deltaopt},\ref{Uopt}) for $U=\{10^{-3},10^{-2},10^{-1},1\}\kappa$ respectively from panel (a) to panel (d). The vertical dashed lines delimit the time window associated with one lifetime $\tau_c=1/\kappa$.}
\label{g2tau_vs_U}
\end{figure}

Finally, to fully characterize the nature of the cavity 1 emission, we compute the delayed second order correlation function in the steady state
\begin{equation}\label{g21tau}
  g_1^{\left( 2 \right)}\left( \tau  \right) = \frac{{\langle {\hat a_1^\dag \left( 0 \right)\hat a_1^\dag \left( \tau  \right){{\hat a}_1}\left( \tau  \right){{\hat a}_1}\left( 0 \right)} \rangle }}{{{{\langle {\hat a_1^\dag \left( 0 \right){{\hat a}_1}\left( 0 \right)} \rangle }^2}}}
\end{equation}
This quantity, involving two-time correlations, is obtained by means of the quantum regression theorem \cite{Gardiner2004,Flayac2017}. We show in Fig.\ref{g2tau_vs_U} the $g_1^{\left( 2 \right)}\left( \tau  \right)$ function for different values of the nonlinearity $U\leq\kappa$. The functions oscillates with a characteristic period $T=\pi/J_{|\rm opt}$  -- which determines an antibunching time window -- and with an amplitude controlled by $\Delta$.

\begin{figure}[ht!]
\includegraphics[width=0.49\textwidth,clip]{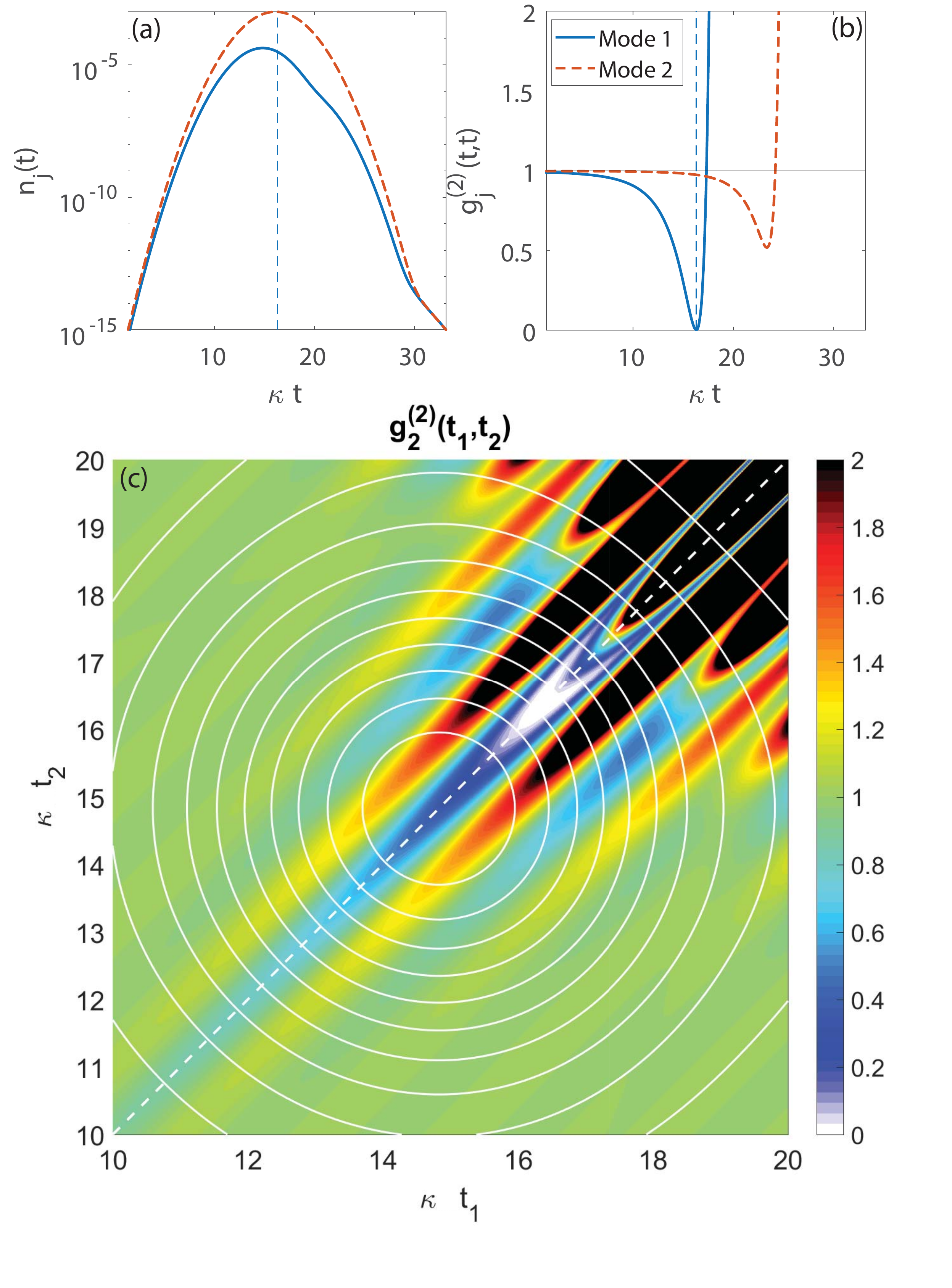}\\
\caption{(Color online) Dynamics of (a) the mean occupancies (log scale) and (b) equal time second order correlation function. The dashed lines mark the position of the $g^{(2)}_1(t,t)$ function. (c) Corresponding two-time second order correlations of the driven cavity $g^{(2)}_1(t_1,t_2)$ under the optimal conditions (\ref{Deltaopt},\ref{Uopt}) for $U=4\times10^{-2}\kappa$ and $f_1=0.1\kappa$. The contours display the two-time occupancy $n_1(t_1,t_2)=\sqrt{n(t_1)n(t_2)}$ and the dashed line stands for the equal time $g^{(2)}_1(t,t)$.}
\label{n_and_g2tt}
\end{figure}

\subsection{Pulsed excitation regime}
In order to use the unconventional photon blockade for single photon applications, the system must be operated under pulsed excitation \cite{Fischer2016}. This requires either suppression of the $g_1^{\left( 2 \right)}\left( \tau  \right)$ oscillations or making them occur on a time scale longer than the cavity lifetime $\tau_c=1/\kappa$. A value of $\Delta=\Delta_{|\rm opt}=0$ is not allowed by Eqs.(\ref{Deltaopt},\ref{Uopt}) as it would require $2J^2=\kappa^2$ and therefore $U_{|\rm opt}\rightarrow\infty$. Besides, as shown in Eq.\eqref{Uopt} and Fig\ref{g2_maps}(b), $U_{|\rm opt}$ increases as $\propto1/J^2$ which imposes a lower bound on $J$ for weakly nonlinear systems where $U\ll\kappa$. For instance, by targeting the limiting case where $\tau_c=\pi/J$ namely $J=\pi\kappa$ results in $U_{|\rm opt}\simeq4\times10^{-2}\kappa$ which is still reasonably weak. However, in practice the bunched parts of the $g_1^{(2)}(\tau)$ function bring additional constraints in the pulsed operation. For illustration, we show in Fig.\ref{n_and_g2tt} the system dynamics following the excitation by a Gaussian pulse $F_1(t)=f_1\exp[-(t-t_0)^2/2\sigma_t^2]$ where $\sigma_t=3/\kappa$ and $t_0=3.5\sigma_t$. Panel (a) shows the cavity occupancy on a semi-log scale and panel (b) the equal-time second-order correlations versus time. The $g^{(2)}_1(t,t)$ function reaches its minimum [dashed-blue line] shortly after the occupancy maximum. In panel (c) we show the two-time second-order correlations $g^{(2)}(t_1,t_2)$ over the time when the pulsed excitation occurs. The plot reveals the oscillations previously discussed for the steady state [see Fig.\ref{g2tau_vs_U}] along the $(t_1,t_1+\tau)$ time axis. The most relevant quantity to study the average emission statistics over a pulse is the second order correlation integrated over two times \cite{Flayac2015}
\begin{equation}\label{g2pulse}
g^{(2)}_{\rm pulse} = \frac{{\int {G_1^{(2)}\left( {{t_1},{t_2}} \right)d{t_1}d{t_2}} }}{{\int {{n_1}({t_1}){n_1}({t_2})d{t_1}d{t_2}} }}
\end{equation}
where $G_1^{(2)}({{t_1},{t_2}})=\langle\hat a_1^\dag(t_1)\hat a_1(t_2)^\dag\hat a_1(t_2)\hat a_1(t_1)\rangle$. In the case we consider here, despite a large antibunching window [white/blue areas], the integrated statistics amounts to $g^{(2)}_{\rm pulse}\approx1.06$ and is therefore classical due to the presence of the bunched regions (red areas) when the occupancy is still sizable. This seems to indicates that the UPB in the terms introduced in Refs.\cite{Liew2010,Bamba2011} cannot be operated under pulsed excitation. One strategy, which was developed in Ref.\cite{Flayac2015}, is to time-gate the output signal in order to specifically target the antibunched regions and extract a nonclassical statistics. For instance, a time window of duration $1/\kappa$ centered on the $g^{(2)}(t,t)$ minimum allows obtaining a value of $g^{(2)}_{\rm pulse}\simeq0.1$ but at the price of an emission rate reduced by a factor of 10. Below, we will see that a mutual driving scheme is in fact sufficient to recover a smooth behavior of the $g^{(2)}(\tau)$ function and therefore a direct compatibility with pulsed excitation.

We note that an alternative strategy to reveal a nonclassical statistics is to take advantage of the coherent population oscillations between the coupled modes after a short excitation pulse. Indeed, as predicted in Ref.\cite{Flayac2017} and measured in Ref.\cite{Adiyatullin2017}, the free evolution of the weakly nonlinear system is accompanied by strong dynamical modulations of the second order correlations. The photon statistics can periodically display sub-Poissonian time windows when the mean occupation oscillates below unity.

\section{Optimal squeezing}\label{Sec:Squeezing}
\subsection{Statistics of a coherent squeezed state}
The unconventional photon blockade can be alternatively described in terms of quadrature squeezing \cite{Lemonde2014} in the limit of weak Kerr nonlinearity where the state remains approximately Gaussian. A \emph{coherent squeezed state} $\hat {\cal D}(\alpha)\hat {\cal S}(\xi)\left| 0 \right\rangle  = \hat {\cal D}(\alpha)\left| \xi  \right\rangle  = \left| {\alpha ,\xi } \right\rangle$ is obtained by the consecutive application of the squeezing $\hat{\cal S}(\xi)=\exp(\xi^*\hat a^2/2-\xi\hat a^{\dag2}/2)$ and displacement $\hat{\cal D}(\alpha)=\exp (\alpha \hat a - {\alpha ^*}\hat a)$ operators respectively defined by the complex parameters $\xi=r\exp(i\theta)$ and $\alpha = \bar \alpha \exp(i \phi)$. The $n$-photon probability distribution ${\cal P}_n = {\left| {\left\langle {n\left| {\alpha ,\xi } \right.} \right\rangle } \right|^2}$ of such a state is given by \cite{Gerry2005}
\begin{eqnarray}
\nonumber  {\cal P}_n &=& \exp\left[{ - \frac{1}{2}\tanh (r)\left( {\alpha^2 {e^{ - i\theta }} + {e^{i\theta }}{\alpha ^{*2}}} \right) - {{\left| \alpha  \label{Pn}
\right|}^2}} \right] \hfill \\
   &\times& \frac{{\tanh^n{{(r)}}}}{{{2^n}\cosh \left( r \right)n!}}{{H}_n}\left[ {\frac{\gamma }{{\sqrt {{e^{i\theta }}\sinh (2r)} }}} \right]
\end{eqnarray}
where we have defined $\gamma\equiv\alpha \cosh(r) + {\alpha ^*}{e^{i\theta }}\sinh(r)$ and ${H}_n$ is the $n^{\rm th}$ Hermite polynomial. In particular the 2-photon probability reads
\begin{equation}\label{P2}
  {\cal P}_2=\frac{1}{8}{\text{sec}}{{\text{h}}^5}\left( r \right){\left[ {\sinh \left( {2r} \right) - 2{{\bar \alpha }^2}{e^{2r}}} \right]^2}{e^{ - \bar \alpha^2 \left[ {1  + \tanh (r)} \right]}}
\end{equation}
\begin{figure}[ht]
\includegraphics[width=0.49\textwidth,clip]{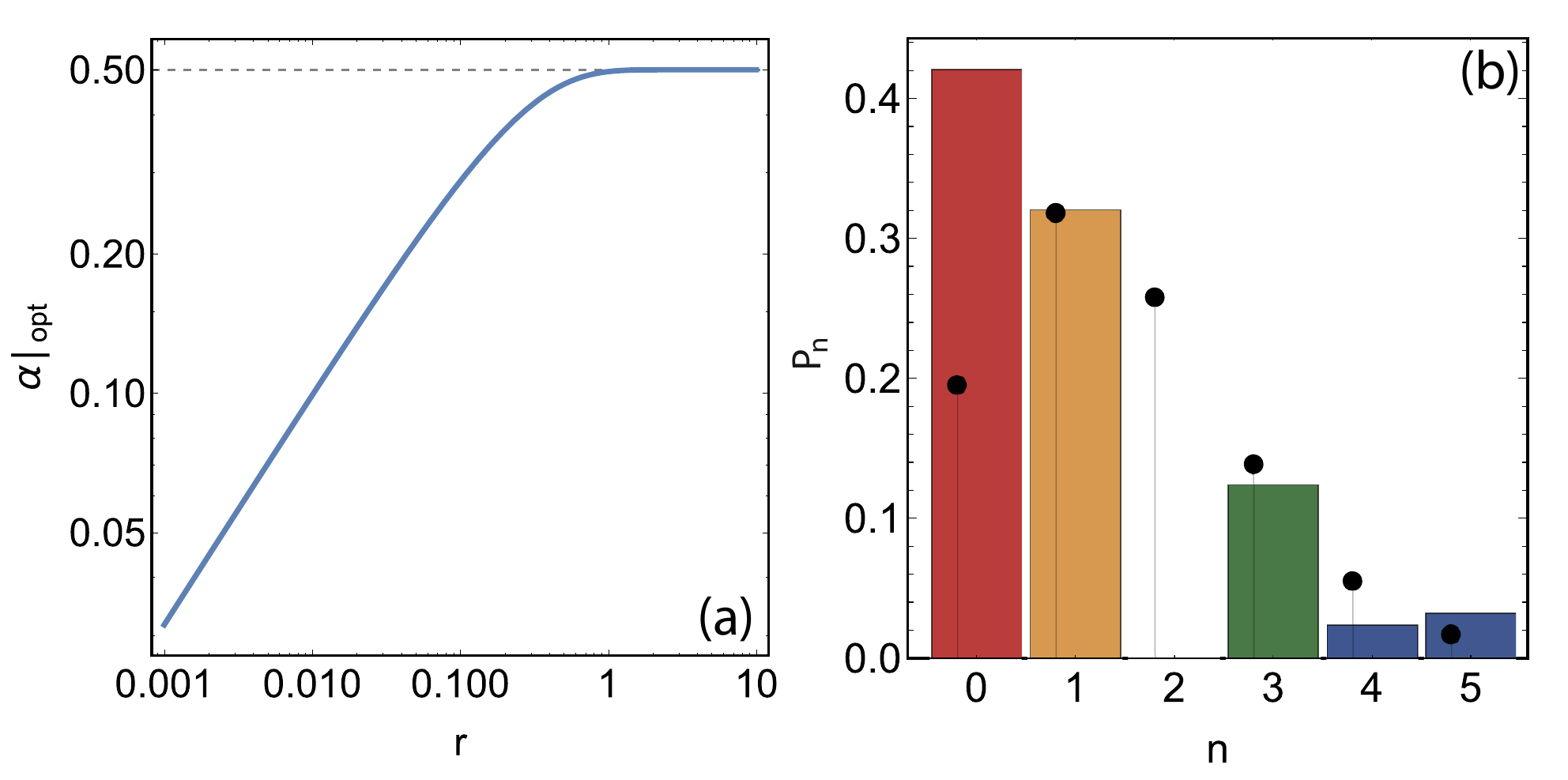}\\
\caption{(a) Optimal displacement $\bar\alpha_{|\rm opt}$ \eqref{alphaopt} as a function of $r$. (b) Probability distribution ${\cal P}_n$ \eqref{Pn} for $r=1$ and $\bar \alpha=\bar\alpha_{|\rm opt}\approx0.5$. The black dots show the corresponding Poissonian distribution for the same average occupation $\bar n$.}
\label{Opt_Displacement}
\end{figure}
We have assumed here the intensity squeezing condition $\theta=2\phi=0$ to favor a sub-Poissonian statistics \cite{Lemonde2014}. One can ask for the conditions for which ${{\cal P}}_2=0$ which results in the relation
\begin{equation}\label{alphaopt}
  \bar\alpha_{|\rm opt}=  \frac{1}{2}{e^{ - 2r}}\sqrt {{e^{4r}} - 1}
\end{equation}
In the limit where $r\rightarrow0$, Eq.\eqref{alphaopt} reduces to $\bar\alpha_{|\rm opt}=\sqrt{r}$ and in the limit $r\rightarrow\infty$, the optimal displacement is bound from above by $\bar\alpha_{|\rm opt}=0.5$ as one can see in Fig.\ref{Opt_Displacement}(a). The average occupation of the coherent squeezed state is $\bar n=|\bar \alpha|^2+\sinh^{2}(r)$ which indicates that a suppression of the 2-photon probability can occur for arbitrarily large photon number if the state is sufficiently squeezed. Figure \ref{Opt_Displacement}(b) shows an example of ${\cal P}_n$ distribution in the optimal squeezing condition for $r=1$.
%We see that while ${\cal P}_2$ is suppressed, the 3-photon  probability ${\cal P}_3$ is enhanced with respect to a Poissonian statistics.

\begin{figure}[ht]
\includegraphics[width=0.49\textwidth,clip]{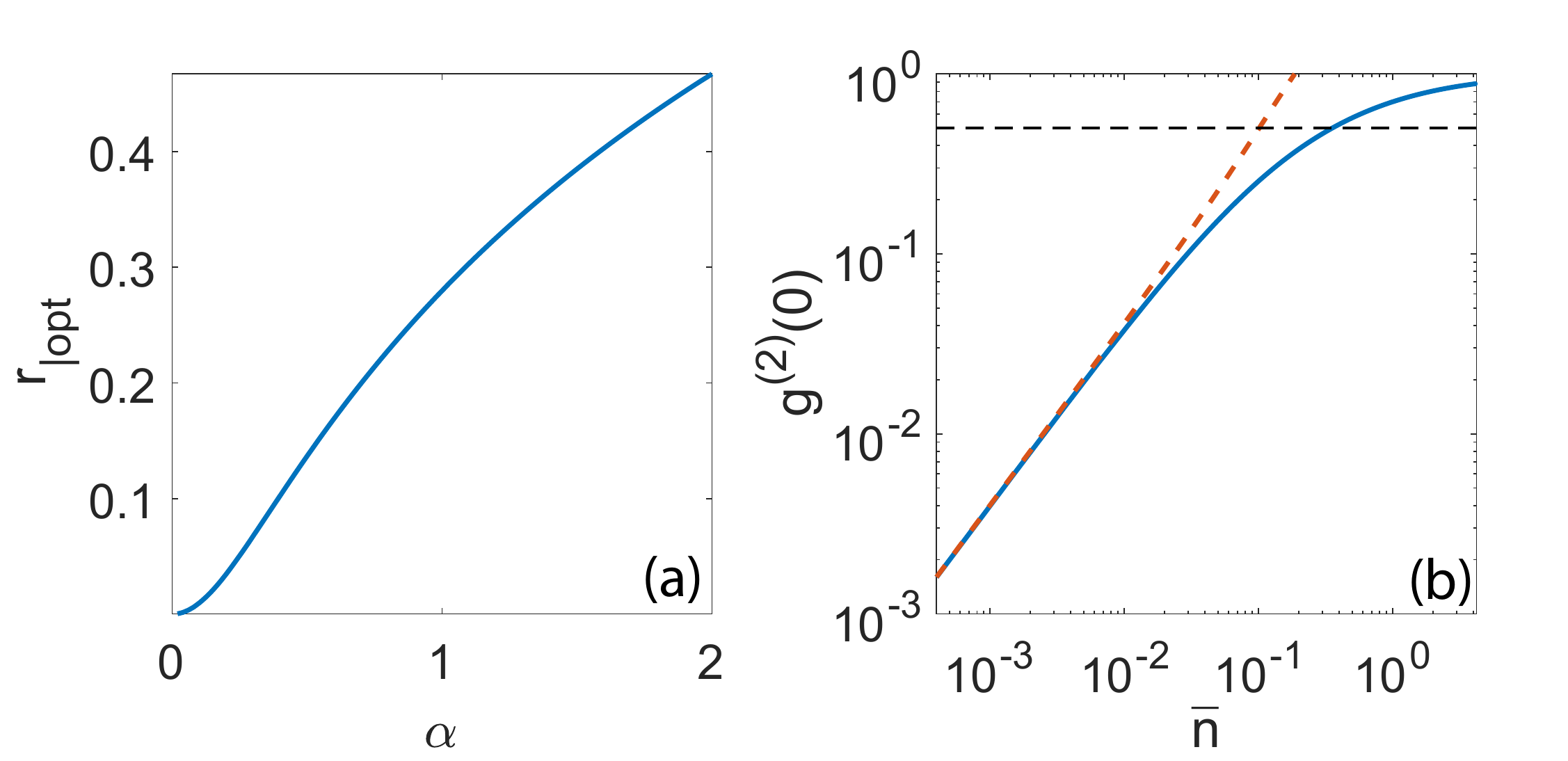}\\
\caption{(a) Optimal value of $r$ versus $\bar\alpha$ in the case for $\theta=2\phi=0$. (b) Second order correlation versus the mean occupancy $\bar n=|\bar \alpha|^2+\sinh^{2}(r)$ (blue line). The dashed-red line shows the values obtained under the condition \eqref{alphaopt}.}
\label{Opt_g2_squeeze}
\end{figure}

The general expression of the second order correlation of the coherent squeezed state is
\begin{equation}\label{g2SC}
  {g^{\left( 2 \right)}}\left( 0 \right) = 1 + \frac{{p^2} + {s^2} + {2\bar \alpha^2 \left[ {p - s\cos \left( {\theta  - 2\varphi } \right)} \right]}}{{{{\left( {{{\bar \alpha }^2} + p} \right)}^2}}}
\end{equation}
with $p=\sinh^2(r)$ and $s=\cosh(r)\sinh(r)$. Sub-Poissonian statistics is indeed favored for $\theta=2\phi$ and one can then minimize Eq.\eqref{g2SC} versus $r$ to obtain the optimal squeezed state for every field amplitude. In the limit $\bar \alpha\rightarrow0$ one simply obtains ${g^{(2)}}(0)\simeq4\bar\alpha_{|\rm{opt}}^2=4 r_{|\rm{opt}}$ which coincides with Eq.\eqref{alphaopt}. The results are summarized in Fig.\ref{Opt_g2_squeeze} where panel (a) shows the optimal $r_{|\rm{opt}}$ value versus $\bar\alpha$ and panel (b) the corresponding values of ${g^{(2)}}(0)$ versus the corresponding mean occupancy $\bar n$, which sets a lower bound for the most general Gaussian state. The dashed-red line corresponds to the condition \eqref{alphaopt}. We see that at low occupancy the two curves are in perfect agreement while, when approaching $\bar n=1$, suppressing ${{\cal{P}}_2}$ becomes sub-optimal as compared to the full ${g^{( 2 )}}(0)$ optimization. A value of ${g^{(2)}}(0)=0.5$ is reached for an occupancy as large as $\bar n\simeq0.35$.

\subsection{Role of squeezing in the UPB}\label{Sec:UPBvsSqueeze}
The degenerate parametric amplifier (DPA) seems to be an obvious candidate for the realization of optimal squeezing \cite{Lemonde2014}. However in practice, it requires a two pump configuration so as to trigger the parametric process from the source mode and to set the suitable displacement of the idler mode. The Kerr nonlinearity is another useful and widely adopted resource for squeezing \cite{Gerry1994,Bajer2002}. It is easily revealed by linearizing the interaction after expanding $\hat a\rightarrow\delta\hat a+\alpha\hat {\mathbb{I}}$ to obtain (up to a displacement and a constant energy shift)
\begin{equation}\label{HK}
{{\hat {\cal H}}_K} = U{{\hat a}^{\dag2} }\hat a^2 \approx U\left( {\alpha^2 {{\delta \hat a}^{\dag 2}} + {\alpha ^{*2}}{{\delta \hat a}^2}} \right)
\end{equation}
Eq.\eqref{HK} is nothing but a DPA interaction of magnitude $\lambda=U|\alpha|^2$ which however binds the subsequent squeezing parameter to the displacement $\alpha$. It therefore prevents the independent tuning of $\alpha$ and $\xi$ required to reach the optimal condition discussed in the previous section. This ultimately illustrates why two coupled Kerr resonators may instead lead to optimal conditions, as the increased number of system parameters allows for independent variation of the displacement and squeezing parameters. Indeed from Eqs.(\ref{rhotf},\ref{Hf}), one can extract the effective parametric interaction \cite{Lemonde2014} seen e.g. by the cavity 1 in the steady state
\begin{equation}\label{lambdaeff}
  \lambda_1^{{\rm{eff}}} = {U_1}\alpha _1^2 - \frac{{{J^2}}}{{{U_2^2}{{\left| {{\alpha _2}} \right|}^4} - {{| {{{\tilde \Delta }_2}} |}^2}}}{U_2}\alpha _2^2\,.
\end{equation}
In the limit where $\lambda_1^{{\rm{eff}}}\ll\kappa_1$, this quantity can be related to a generic squeezing parameter $\xi_1$ as $r_1\simeq2|\lambda _1^{\rm eff}|/\kappa$ and $\theta_1=\arg(\lambda_1^{{\rm{eff}}})$. The latter can also be directly computed from the quantum fields \cite{Bajer2002} following
\begin{eqnarray}
\label{rexpr}
  {r_j} &=& {\left| \langle {\hat a_j^2} \rangle  - {\langle {{{\hat a}_j}} \rangle ^2} \right| + \left| {{{\langle {{{\hat a}_j}} \rangle }}} \right|^2 - \langle {\hat a_j^\dag {{\hat a}_j}} \rangle} \hfill \\
  {\theta_j} &=& \arg\left[{\langle {\hat a_j^2} \rangle  - {\langle {{{\hat a}_j}} \rangle ^2}}\right]
\end{eqnarray}

\begin{figure}[ht!]
\includegraphics[width=0.49\textwidth,clip]{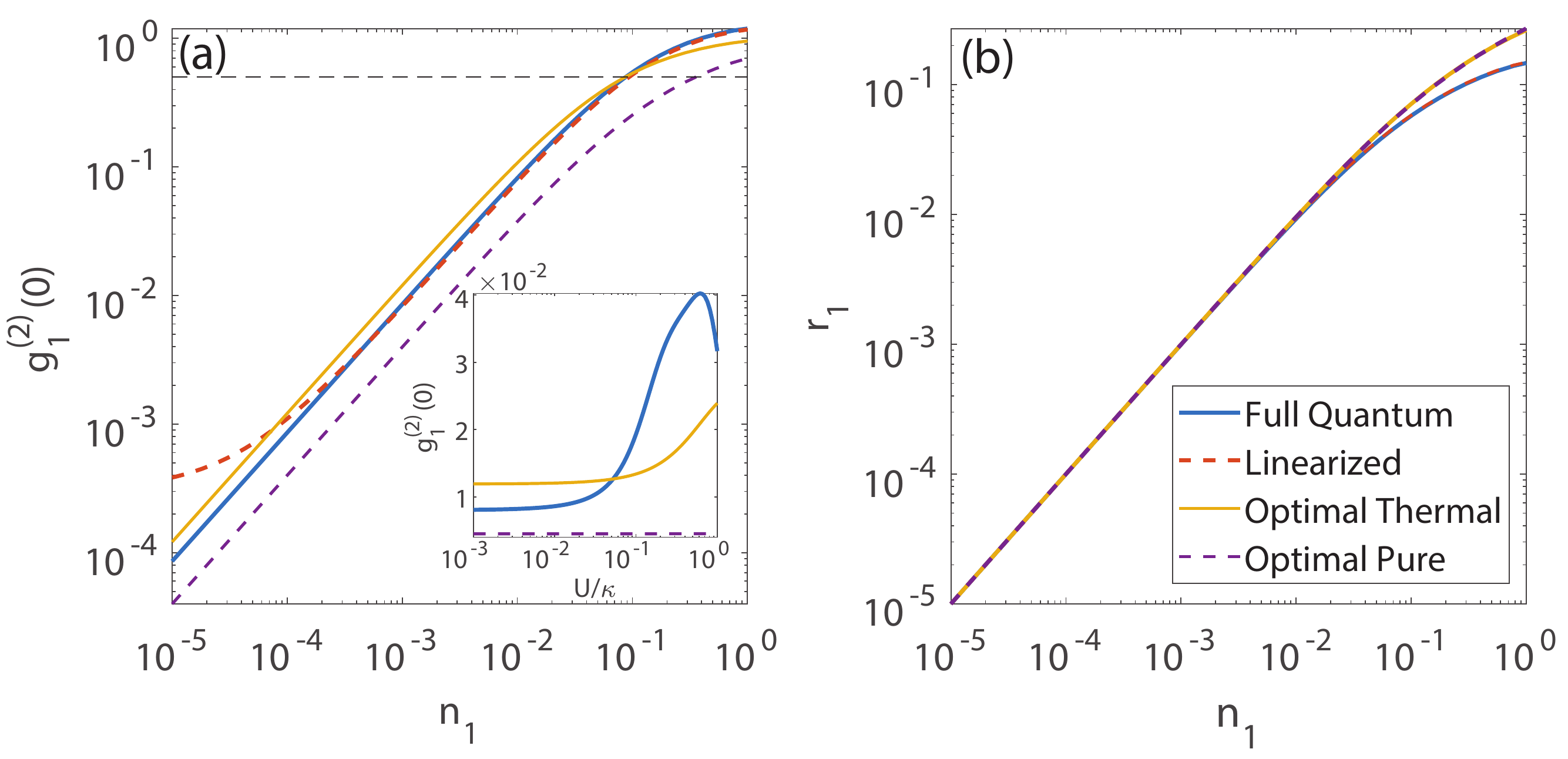}\\
\caption{Comparison between UPB and the optimal squeezed state in the case where $U=10^{-2}\kappa$. (a) Second order correlation function. The inset shows the variations of $g_1^{(2)}(0)$ versus the nonlinearity $U$. (b) Corresponding squeezing parameter versus the mean occupancy $n_1$.}
\label{UPB_vs_Opt}
\end{figure}

Drive and dissipation unavoidably induce some degree of mixdness of the state, which can be quantified by the purity of the density matrix $P={\rm Tr}(\hat\rho^2)$. One can then link the mixdness to an effective thermal noise with average occupation $\bar n_{\rm eff}=(1-P)/(2P)$ \cite{Lemonde2014}. This allows comparing the UPB states with the most general form of Gaussian state namely a \emph{thermal squeezed coherent state}. The corresponding second order correlation function is obtained from Eq.\eqref{g2SC} with $s=(\bar n_{\rm eff}+1/2)\cosh(2r)-1/2$ and $p=(\bar n_{\rm eff}+1/2)\sinh(2r)$ and the mean occupation is $\bar n=\bar \alpha^2+p$.
We show in the figure \ref{UPB_vs_Opt} a comparison between the UPB and the corresponding optimally squeezed state both in the pure and thermal cases for $U=10^{-2}\kappa$. We also display the linearized results obtained by neglecting the second line of Eq.\eqref{Hf} which leads to purely Gaussian states.
While the optimal squeezing is achieved in all cases, as seen in panel (b) showing $r_1$ versus $n_1$, the second order correlation curves display a clear hierarchy. We see that while the UPB (blue line) obviously stands above the pure-state limit (yellow line) due to mixdness, it remarkably lies below the thermal limit (dashed-purple line). This feature cannot be attributed to a possible non-Gaussian nature of the state, as the linearized result (dashed-red line) is in very good agreement with the full quantum one, suggesting a close-to-Gaussian state. The actual explanation resides in the fact that the UPB state presents a form of mixdness far from that of a thermal state. In the inset of panel (a), we show the impact of the nonlinearity for a fixed occupancy $n_1=10^{-3}$. With increasing nonlinearity, the non-Gaussian character results in an increasing second order correlation function that crosses the thermal limit at $U=5\times10^{-2}\kappa$. The subsequent drop of $g_1^{(2)}(0)$ when approaching $U=\kappa$ is a signature of the onset of the standard blockade mechanism.
\section{UPB for arbitrary system parameters}\label{Sec:Developments}
In this section, we shall report on a generalized scheme for UPB. In particular we will show that it is possible to completely relax the link between the intrinsic system parameters $\Delta$, $J$, $U$ and $\kappa$, required in the original proposal for optimal UPB. Optimal conditions can be instead achieved by driving both modes with the proper relative phase and amplitude, which are given by compact analytical formulas. This finding indicates that, in an experimental realization of UPB on a given photonic platform, fine tuning of the intrinsic system parameters is not strictly needed. Moreover, it shows that UPB can be achieved in a two-resonator scheme with small mode coupling $J$ and small detuning $\Delta$, which should enable operability under pulsed excitation. More generally, the input-output theory will reveal the possibility to work with completely uncoupled optical modes upon an adequate mixing of their outputs. We will then discuss alternative implementations of the UPB in a weakly coupled Jaynes-Cummings or optomechanical system to show the universality of the mechanism.

\subsection{Dissipative coupling}\label{Sec:Dissipative}
\begin{figure}[ht!]
\includegraphics[width=0.49\textwidth,clip]{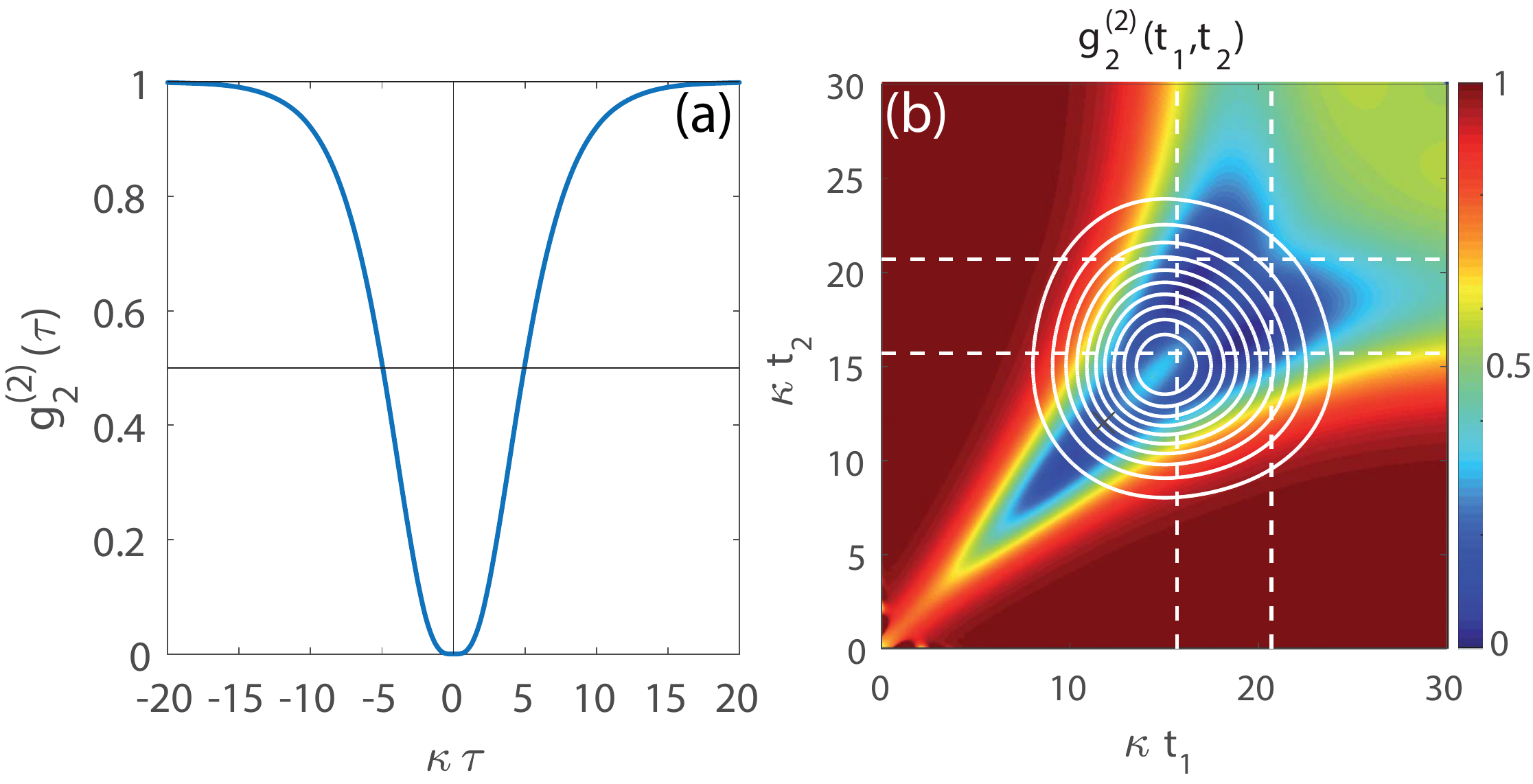}\\
\caption{(a) Delayed second order correlation in the steady state for $\kappa_1=\kappa_2=\kappa$, $U=10^{-2}\kappa$, $\Delta_{1,2}=0$, $F_2=10^{-2}\kappa$, $F_1={F_{1|{\rm{opt}}}}$ and $\chi=\kappa$. (b) Two time second order correlation for a Gaussian pulsed excitation of duration $5/\kappa$. The white contour shows the occupancy $n_1(t_1,t_2)=\sqrt{n(t_1)n(t_2)}$.}
\label{Fig:g2_diss}
\end{figure}

The interpretation of the UPB in terms of optimal squeezing, discussed in Sec.\ref{Sec:UPBvsSqueeze}, suggests that UPB may be achieved in a scheme where the input is processed in two subsequent stages: One producing the squeezing and the following one displacing the resulting field appropriately. Recently there was a growing interest in nonreciprocal photonic structures in view of creating optical isolation or topological states of light \cite{Haldane2008,Yu2009,Feng2011,Hafezi2013,Bernier2016}. It naturally led us to investigate the case of a dissipative interaction between the cavities \cite{Flayac2016} instead of a coherent hopping. We notice that, differently from the original proposal then, here the field displaying UPB will be that of the second (i.e. target) cavity. A unidirectional transmission between two quantum modes is treated within the formalism of cascaded quantum systems \cite{Gardiner1993,Carmichael1993}. If the output of cavity 1 (source) is driving the input of the cavity 2 (target) then the corresponding Linblad term to add in Eqs.(\ref{rhot},\ref{rhotf}) reads
\begin{equation}\label{Da1a2}
i\chi\hat{\cal{D}}\left[ {{{\hat a}_1},{{\hat a}_2}} \right]\hat \rho = i\chi([ {{{\hat a}_1}\hat \rho ,\hat a_2^\dag } ] + [ {{{\hat a}_2},\hat \rho \hat a_1^\dag } ])
\end{equation}
where $\chi=\sqrt{\eta\kappa_1\kappa_2}$ and $\eta\in[0,1]$ is a measure of the one-directional coupling efficiency. The analytical formalism of Sec.\ref{Sec:WeakDriving} can still be applied, in the spirit of Ref.\cite{Carmichael1993}, by setting $J=0$ and by adding the non-Hermitian jump operator $i\chi\hat a_2^\dag\hat a_1$ to Eq.\eqref{HNH}. It then simply translates into an effective non-reciprocal hopping term. For such a scheme, it is crucial to drive both modes namely to have $F_{1,2}\neq0$. Indeed, the target cavity is fed by the squeezed output of the source while the laser will set the right amount of displacement to reach the optimal squeezing condition or equivalently the quantum interference. In the absence of driving of the target cavity, the latter would behave as a bare filter which is not sufficient to produce sub-Poissonian light. As before, an optimal condition can be derived by solving Eqs.(\ref{c10}-\ref{c11}) and requiring that $c_{02}$ vanish. The advantage however is that this condition can now be achieved as a relation between the \emph{complex} driving amplitudes $F_{1}$ and $F_{2}$ for any given values of the intrinsic system parameters. The resulting equation reads
\begin{equation}\label{F1opt}
{F_{1|{\rm{opt}}}} = i{F_2}\frac{{\tilde \Delta {{\tilde U}_1} \pm \sqrt {{U_1}{{\tilde U}_1}\tilde \Delta {{\tilde \Delta }_2}} }}{{\left( {\tilde \Delta  + {U_1}} \right)\chi }}
\end{equation}
with the definitions $\tilde \Delta  \equiv {{\tilde \Delta }_1} + {{\tilde \Delta }_2}$, ${{\tilde U}_1} \equiv {{\tilde \Delta }_1} + {U_1}$ and assuming $F_2\in {\mathbb{R}^+}$ without loss of generality. The condition \eqref{F1opt} doesn't depend on $U_2$ meaning the target cavity could be a purely harmonic mode as it is the case for the driven mode of the original scheme \cite{Bamba2011}. Besides, an optimal condition can be found for vanishing detunings $\Delta_{1,2}$. This results in the delayed second order correlation function of the target cavity varying smoothly instead of oscillating, as shown in Fig.\ref{Fig:g2_diss}(a). As an additional advantage, when comparing this result with the result of Fig.\ref{g2tau_vs_U}, the antibunching time window extends over several lifetimes in the present case. The single photon regime $g^{(2)}_2(\tau)<0.5$ is ensured over at least 5 lifetimes. Panel (b) displays the corresponding two-time correlation map under pulsed excitation. The photon statistics over a pulse, computed from Eq.\eqref{g2pulse}, gives $g^{(2)}_{\rm pulse}\simeq0.3$ which can be reduced below 0.1 by additionally time-gating the output pulse \cite{Flayac2015} over a time window of duration $\Delta T=5/\kappa$ delimited by the dashed-white lines in the plot.

The source cavity behaves as a squeezed source for the target cavity. We deduce that the cascaded configuration described above is equivalent to that of a single cavity driven by a displaced squeezed vacuum as studied in Ref.\cite{Lemonde2014}. This can be modeled by considering a bare driven dissipative mode ${\hat{\cal{H}}}=\Delta \hat a^{\dag}\hat a + F\hat a^{\dag}+F^{*}\hat a$ whose coupling to a squeezed vacuum reservoir is introduced in the master equation as \cite{Dum1992}
\begin{eqnarray}\label{rho_svac}
i\frac{{\partial \hat \rho }}{{\partial t}} &=& \left[ {{\hat{\cal{H}}},\hat \rho } \right] - i\frac{\kappa }{2}{\hat{\cal{D}}}\left[ {\hat a} \right]\hat \rho \\
\nonumber  &+& i\frac{\kappa }{2}{\xi ^*}{\hat{\cal{D}}}\left[ {{{\hat a}^2}} \right]\hat \rho  + i\frac{\kappa }{2}\xi {\hat{\cal{D}}}\left[ {{{\hat a}^{\dag 2}}} \right]\hat \rho
\end{eqnarray}
where here ${\hat{\cal{D}}}\left[ {{{\hat o}^{2}}} \right]\hat \rho=\{{{\hat o^2}},\hat \rho\} - 2{{\hat o}}\hat \rho \hat o$ and $\xi=r\exp{i\theta}$ is the squeezing parameter. On can show that, for small occupations, the optimal squeezing condition is achieved for $r\simeq|\langle\hat a\rangle|^2$ and $\theta=2\arg{\langle\hat a\rangle}$ as expected. It can be reached by tuning the amplitude and phase of the driving field $F$ for a given value of $\xi$. The cascaded cavity configuration can therefore be pictured as a system where the squeezed source is directly integrated in the structure.

\subsection{Input-Output theory}\label{Sec:IOTheory}
\subsubsection{Optimal driving fields}
\begin{figure}[ht!]
\includegraphics[width=0.3\textwidth,clip]{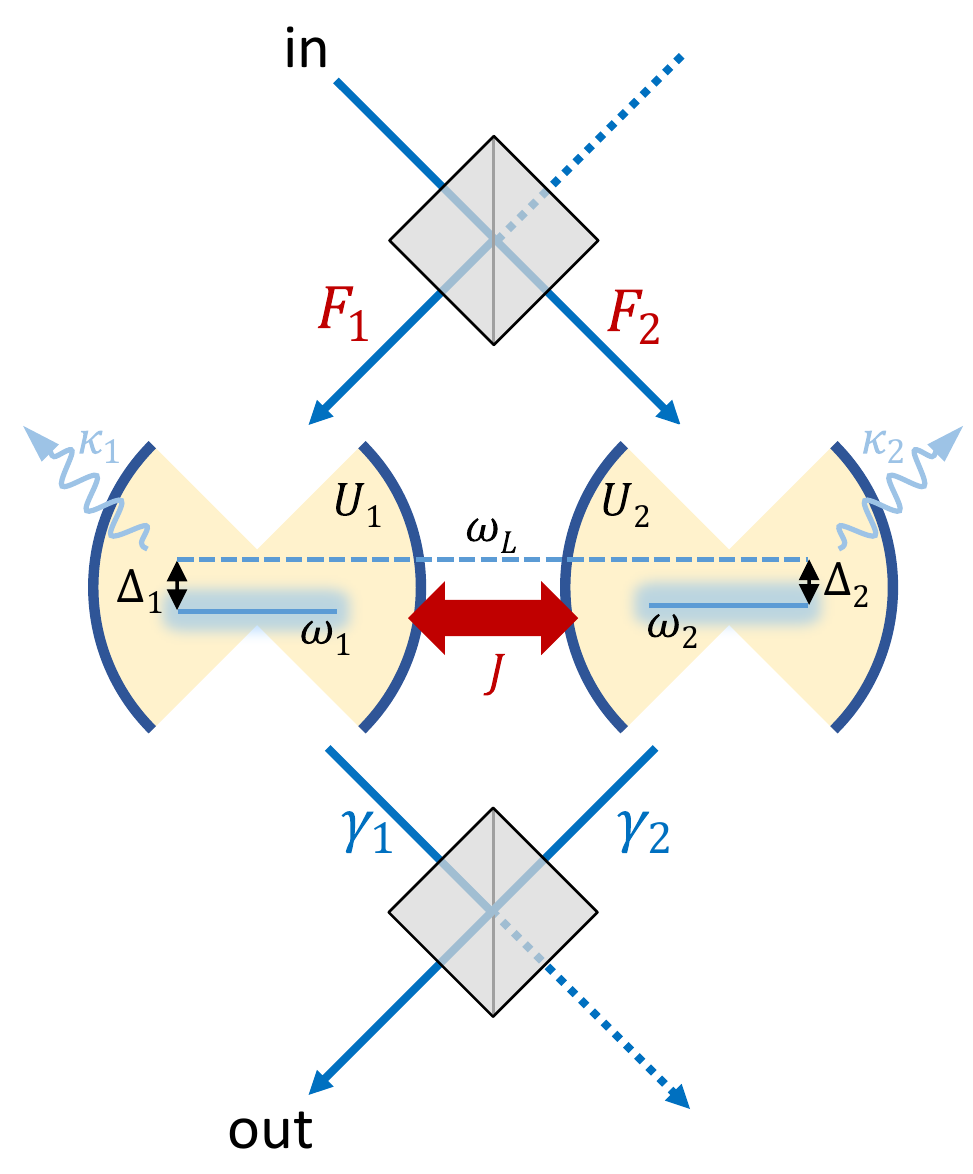}\\
\caption{Scheme of the input-output mixing scheme: each cavity mode is driven by a mutually coherent field of complex amplitude $F_{1,2}$ and the output fields are mixed in proportions set by the complex coefficients $\gamma_{1,2}$.}
\label{Fig:Mixing}
\end{figure}

A natural question that can be asked at that stage is, whether the optimization strategy with two driving fields of Sec.\ref{Sec:Dissipative} can be applied also to the original case with coherent coupling. This scheme was studied Refs.\cite{Flayac2013,Shen2015,Wang2017,Yu2017} where the authors derived some optimal values of the nonlinearity and detuning in the presence of a bilateral drive. Here we follow the most natural approach of Sec.\ref{Sec:Dissipative} and solve Eqs.(\ref{c10},\ref{c11}) in the case where $J\neq0$ and $F_{1,2}\neq0$ [see Fig.\ref{Fig:Mixing}]. We obtain the optimal condition
\begin{equation}\label{F1opt_coh}
{F_1}{|_{\rm opt}} = \frac{{{F_2}\tilde \Delta {{\tilde U}_1}J \pm \sqrt {F_2^2{J^2}{U_1}\left[ {{{\tilde \Delta }_2}\tilde \Delta {{\tilde U}_1} - {J^2}\left( {\tilde \Delta  + {U_1}} \right)} \right]} }}{{{J^2}\left( {\tilde \Delta  + {U_1}} \right)}}\,.
\end{equation}
absorbing all the parameter constraints in the driving fields. This again shows that, even for the coherent coupling case, arbitrary values of the intrinsic system parameters can be assumed, provided the driving fields are appropriately tuned in amplitude and phase. In particular, not only ar we able to consider any value of $\Delta_{1,2}$, but also arbitrarily small values of the coupling which can e.g. be set to $J<\kappa$ so as to suppress the oscillations of the $g^{(2)}(\tau)$ function.

\subsubsection{Output Mixing}
Let us now study the opposite situation in which arbitrary driving fields are present and the output fields are instead mixed as sketched in Fig.\ref{Fig:Mixing}. The standard input-output theory \cite{Gardiner1993} allows us to write the global output field as
\begin{equation}\label{aout}
  \hat a_{\rm out} = \hat a_{\rm in} + \gamma_1 \hat a_1 + \gamma_2 \hat a_2
\end{equation}
where $\hat a_{\rm in}$ is an input noise operator. For zero thermal occupations $\bar n_{\rm th}=0$ all the normally ordered field/noise correlations vanish \cite{Gardiner2004} and the subsequent output occupation and second order correlation read \cite{Flayac2013}
\begin{eqnarray}\label{g2out}
{n_{\rm out}} &=& \langle {\hat a_{\rm out}^\dag {{\hat a}_{\rm out}}} \rangle \approx {\left| {\gamma _1^2{c_{10}} + \gamma _2^2{c_{01}}} \right|^2} \hfill \\
g_{{\rm{out}}}^{(2)}(0) &=& \frac{{\langle \hat a_{\rm out}^\dag \hat a_{\rm out}^\dag {{\hat a}_{\rm out}}{{\hat a}_{\rm out}}\rangle }}{{n_{\rm out}^2}} \hfill \\
\nonumber                 &\simeq& \frac{{{{\left| {\gamma _1^2{c_{20}} + {\gamma _1}{\gamma _2}\sqrt 2 {c_{11}}} \right|}^2} + {{\left| {\gamma _2^2{c_{02}} + {\gamma _1}{\gamma _2}\sqrt 2 {c_{11}}} \right|}^2}}}{{n_{\rm out}^2}} \hfill \\
                          &+& \frac{{{{\left| {\gamma _1^2{c_{20}} + \gamma _2^2{c_{02}}} \right|}^2}}}{{n_{\rm out}^2}}
\end{eqnarray}
We can now look for the values of the coefficients $\gamma_{1,2}$ that realize the condition $g_{{\rm{out}}}^{(2)}(0)\simeq0$. Remarkably, provided that both cavities are driven, one can consider the fully decoupled case for which $J=0$. Assuming further identical cavities by setting $U_1=U_2=U$, $\Delta_1=\Delta_2=\Delta$ and $\kappa_1=\kappa_2$, we obtain the optimal output condition
\begin{equation}\label{gamma1opt}
  {\gamma _1}|_{\rm opt} = {\gamma _2}\frac{{\sqrt {F_1^2F_2^2\left( {2\tilde \Delta  + U} \right)U}  \pm {F_1}{F_2}\left( {\tilde \Delta  + U} \right)}}{{F_1^2\tilde \Delta }}
\end{equation}

In view of the realization of a single-photon source, it is interesting to look for the parameters required to allow for a perfectly symmetric input $F_1=F_2=F$ or output $\gamma_1=\gamma_2=\gamma$ where $\{F,\gamma\}\in{\mathbb R}^+$. Under these requirements we obtain
\begin{eqnarray}
    {\gamma _1}{|_{{\rm{opt}}}} &=& {\gamma _2}\frac{{\sqrt {\left( {2\tilde\Delta  + U} \right)U}  \pm \left( {\tilde\Delta  + U} \right)}}{\tilde\Delta } \hfill \\
    {F_1}{|_{{\rm{opt}}}} &=& {F_2}\frac{{\sqrt {\left( {2\tilde\Delta  + U} \right)U}  \pm \left( {\tilde\Delta  + U} \right)}}{\tilde\Delta}\,,
\end{eqnarray}
assuming $F_2\in{\mathbb R}^+$ and $\gamma_2\in{\mathbb R}^+$. Hence, in the weak driving limit $F_{1,2} \rightarrow 0$, two separate optimal conditions hold for the input and output parameters respectively. We note that there is no condition for which the system is fully symmetric namely $F_1=F_2$ and $\gamma_1=\gamma_2$ even by allowing distinct parameters for the two cavity modes.

\begin{figure}[ht]
\includegraphics[width=0.5\textwidth,clip]{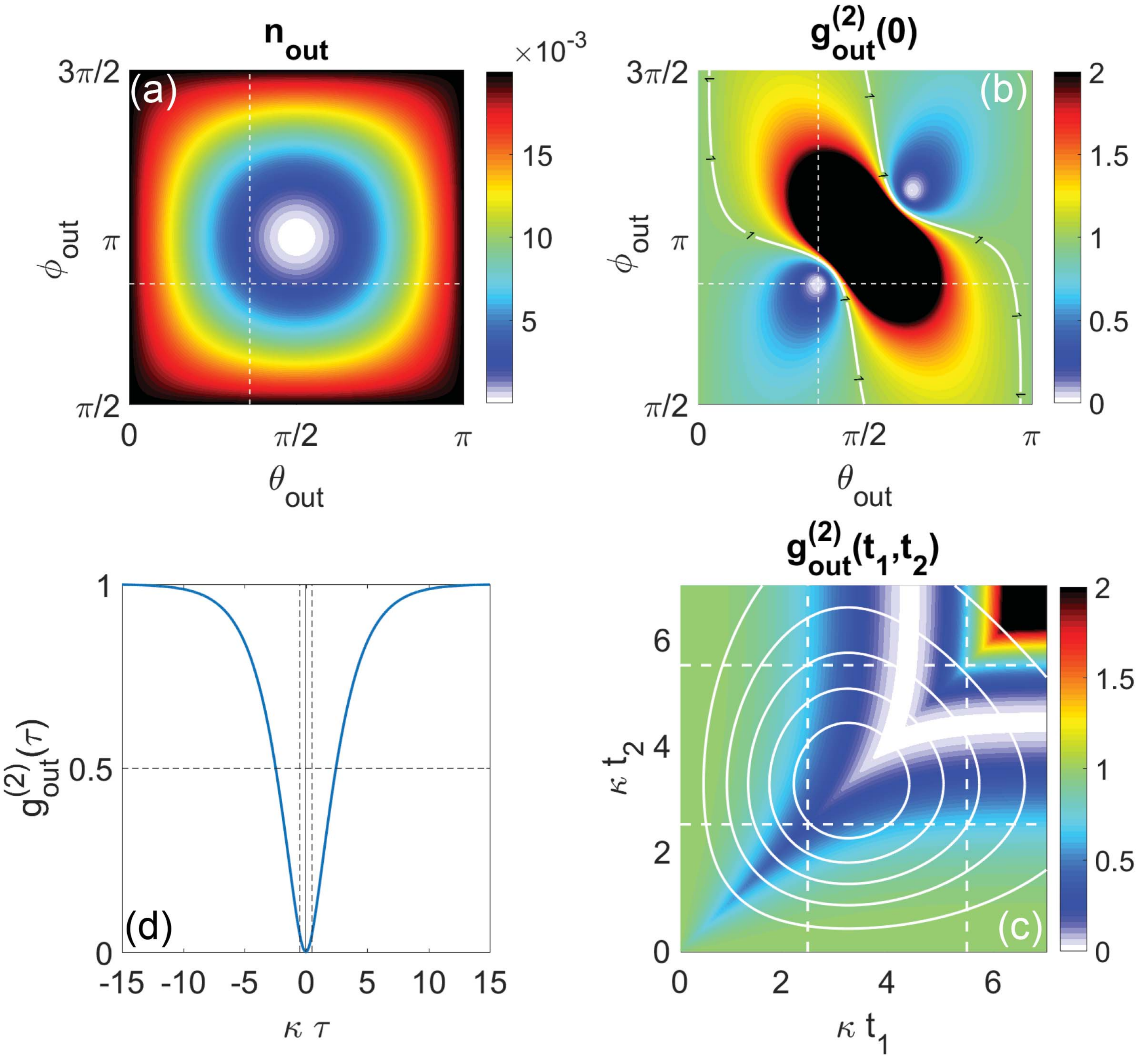}\\
\caption{(Color online) Maps of the output (a) population $n_{\rm out}$ and (b) second order correlation $g^{(2)}_{\rm out}(0)$ as a function of $\theta_{\rm out}$ and $\phi_{\rm out}$. The parameters are $U=10^{-2}\kappa$, $J=0$, $F_0=10^{-1}\kappa$, $\theta_{\rm in}=\pi/2$, $\phi_{\rm in}=0$, $\Delta_{1,2}=0$. (c) Delayed second order correlations $g^{(2)}_{\rm out}(\tau)$ at a minimum of the panel (b) map (dashed-white lines) corresponding to the optimal output condition. (d) Two-time second order correlation map under pulsed excitation.}
\label{Fig:Output}
\end{figure}

To study the occurrence of antibunching as a function of input and output parameters, it is convenient to define these parameters in the Stokes representation as
\begin{eqnarray}
\label{Fin}
  {F_1} &=& F_0\cos \left( {{\theta _{{\rm{in}}}}/2} \right), \hfill {F_2} = F_0\sin \left( {{\theta _{{\rm{in}}}}/2} \right){e^{i{\varphi _{{\rm{in}}}}}} \\
\label{gammaout}
  {\gamma _1} &=& \gamma_0\cos \left( {{\theta _{{\rm out}}/2}} \right), \hfill {\gamma _2} = \gamma_0\sin \left( {{\theta _{{\rm{out}}}}/2} \right){e^{i{\varphi _{{\rm{out}}}}}}
\end{eqnarray}
where $\theta_{\rm{in,out}}$ control the relative amplitudes and $\phi_{\rm{in,out}}$ the relative phases. For given values of $F_0 = 10^{-1}\kappa$, $\theta_{\rm in}=\pi/2$, $\phi_{\rm in}=0$ (equal driving), and assuming a perfect detection namely $\gamma_0=\sqrt{\kappa}$, we plot in Fig.\ref{Fig:Output}(a),(b) the $n_{\rm out}$ and $g_{{\rm{out}}}^{(2)}(0)$ maps obtained by varying $\theta_{\rm out}$ and $\phi_{\rm out}$ in the cases $J=0$. Strong antibunching areas appear in the white/blue regions. In Fig.\ref{Fig:Output}(c), we show the delayed second order correlations $g^{(2)}_{\rm out}(\tau)$ computed at a minimum of the function in panel (b), corresponding to the optimal output condition \eqref{gamma1opt}. Finally in Fig.\ref{Fig:Output}(d), we show the two-time second order correlation map obtained from a pulsed excitation simulation. We used here a pulse of duration $\sigma_t=1/\kappa$ with the very same parameters as previously. We obtain an integrated value -- as defined in Eq.\eqref{g2pulse} -- of $g^{(2)}_{\rm pulse}\simeq0.4$ which drops e.g. to $0.1$ when the time window delimited by the dashed lines is targeted \cite{Flayac2015}. These values could be greatly improved by optimizing the pulse duration and/or its temporal shape.

The assumptions leading to the results shown in Fig.\ref{Fig:Output} are particularly well suited to model the case of a single cavity with two degenerate resonant modes of orthogonal polarization \cite{Bamba2011a}, driven by a laser polarized according to Eq.\eqref{Fin} and a suitably selected polarization angle for the detection, defined by Eq.\eqref{gammaout}. Systems with these features are those based on a semiconductor micropillar etched out of a planar semiconductor microcavity with distributed Bragg reflectors \cite{Shelykh2010a}. The nonlinearity can be implemented through an embedded quantum well -- whereby coupling to the excitons results in microcavity polaritons \cite{Boulier2014} -- or by including one semiconductor quantum dot \cite{Gazzano2013} -- resulting in a physical realization of the Jaynes-Cummings model as discussed in the next Section. Progress in terms of fabrication quality and photon lifetime for these systems has been remarkable in the last decade \cite{Sun2016}, and very recently strongly bunched photon statistics has been experimentally demonstrated \cite{Adiyatullin2017}.

We conclude this paragraph by pointing out that the case where e.g $U_1\neq0$, $U_2=0$ and $J=0$ models a configuration close to the proposal of Ref.\cite{Kitagawa1986}. In that work, the output of a Kerr oscillator is mixed with the input through a delay line setting the suitable displacement to produce antibunching in the output field. This scheme opens the way to integrated single photon emission from a bare Kerr resonator. It could be easily implemented with photonic crystal cavities \cite{Xu2014a} to realize a ``self-homodyning'' scheme \cite{Fischer2017}. Indeed, recent progress in design optimization has produced photonic crystal cavities displaying an ultra-high quality factor, both in silicon \cite{Lai2014} and in wide-bandgap materials \cite{VicoTrivino2014}. These latter have demonstrated high-efficiency optical nonlinearity of both second and third order \cite{Mohamed2017}.

\subsection{Alternative Systems}\label{Sec:Alternative}
The UPB can be realized in many different configurations involving Kerr nonlinearities, which may be a route to the realization of a passive single-photon source \cite{Ferretti2013,Flayac2015}. UPB can however be obtained in the presence of other kinds of nonlinearities. The first example is that of the Jaynes-Cummings model in the so-called dispersive regime, where the detuning $\Delta_{ce}=\omega_c-\omega_e$ between the cavity and the two level emitter is much larger than their mutual coupling $g$. In this limit, the Jaynes-Cummings modes results in an effective Kerr nonlinearity $U_{\rm eff}=g^4/\Delta_{ce}^3$ once the two-level system has been traced out \cite{Savage1990,Boissonneault2009}. An equivalent configuration to that of Sec.\ref{Sec:OriginalUPB} would then be that of two coupled cavities where at least one of them hosts a two level emitter \cite{Bamba2011,Cheng2017}. In this case one can easily recover the optimal UPB condition for the effective Kerr nonlinearity $U_{\rm eff}$, which would give rise to a sub-Poissonian statistic even in the weak coupling regime $g\ll\kappa$.

The optomechanical interaction \cite{Aspelmeyer2014} ${\hat{\cal H}}_{\rm om}=g{{\hat a}^\dag }\hat a\hat x$, where $\hat x$ is the position operator of a mechanical oscillator with resonant frequency $\Omega_m$, can also be mapped to a Kerr nonlinearity $U_{\rm eff}=g^2/\Omega_m$ via a polaron transformation \cite{Rabl2011}. It allows realization of the UPB \cite{Xu2013,Savona2013} in such a hybrid system where the typical regime of parameter $g\ll\kappa$ normally prevents the conventional blockade from occurring.

\begin{figure}[ht]
\includegraphics[width=0.49\textwidth,clip]{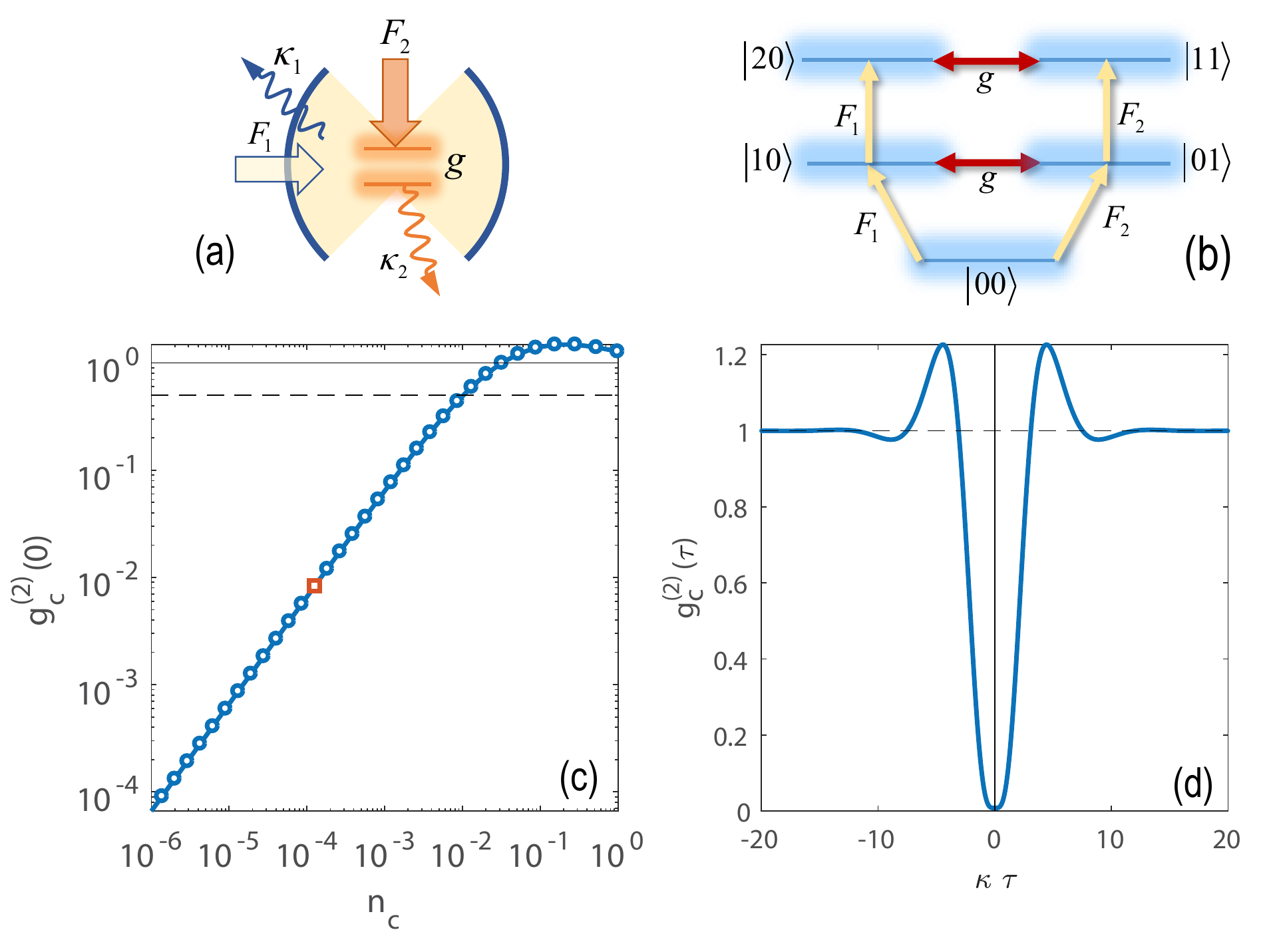}\\
\caption{(Color online) (a) Scheme of the cavity QED system. (b) Bare energy levels in the 2-photon manifold. (c) Second order correlation function of the cavity $g^{(2)}_c(0)$ versus its occupation $n_c$. The parameters are $\Delta_2=0$, $\Delta_1={\Delta _{1|{\rm{opt}}}}=0$, $\kappa_1=\kappa_2=\kappa$, $g={g_{|{\rm{opt}}}}=\kappa/\sqrt{2}$. (d) Two-time second order correlations computed for the red square of panel (c).}
\label{Fig:g2opt_JC}
\end{figure}

Beyond the Kerr nonlinearity, the UPB was also shown to be achievable with second order $\chi^{(2)}$ nonlinearity \cite{Gerace2014,Zhou2015} but also very recently in the framework of the driven-dissipative Rabi model for an arbitrarily strong coupling \cite{Deng2017} between a cavity and an emitter. Here we shall present the simplest configuration possible, sketched in Fig.\ref{Fig:g2opt_JC}(a), allowing exploration of the UPB in the weak coupling regime $g\ll\kappa$ of a cavity QED system. We consider the original Jaynes-Cummings Hamiltonian in the frame rotating at the cavity frequency
\begin{eqnarray}
\nonumber {\cal \hat H} &=& {\Delta _1}{{\hat a}^\dag }\hat a + \frac{\Delta _2}{2}{{\hat \sigma }_ + }{{\hat \sigma }_ - } \hfill + g\left(\hat a^{\dag}\hat\sigma_- + \hat\sigma_+\hat a\right)\\
 &+& {F_1}{{\hat a}^\dag } + F_1^*\hat a + {F_2}{{\hat \sigma }_ + } + F_2^*{{\hat \sigma }_ - }
\end{eqnarray}
where we allow the emitter to be directly driven. The master equation is obtained from Eq.\eqref{rhot} via the substitution $\hat a_2\rightarrow\hat \sigma_-$, $\hat a_2^\dag\rightarrow\hat \sigma_+$ and $\Delta_2\rightarrow\Delta_2/2$. The Sec.\ref{Sec:WeakDriving} treatment can be directly applied simply by disregarding the $c_{02}$ coefficient in the expansion leading to Eq.\eqref{psit}. In the case where $F_2=0$, there exists an optimal UPB condition resulting in a suppression of the 2 cavity photon probability $|c_{20}|^2$ requiring
\begin{eqnarray}
\label{D1opt_JC}
  {\Delta _{1|{\rm{opt}}}} &=&  - \frac{{{\Delta _2}\left( {{\kappa _1} + 2{\kappa _2}} \right)}}{{2{\kappa _2}}} \hfill \\
\label{gopt_JC}
  {g_{|{\rm{opt}}}} &=& \pm \frac{{\sqrt {\left( {\Delta _2^2 + \kappa _2^2} \right)\left( {{\kappa _1} + {\kappa _2}} \right)} }}{{2\sqrt {{\kappa _2}} }}
\end{eqnarray}
which for the resonant case $\Delta_1=\Delta_2$ with equal losses $\kappa_1=\kappa_2$ simplifies to ${g_{|{\rm{opt}}}}=\kappa/\sqrt{2}$ and ${\Delta _{1|{\rm{opt}}}}=0$. This result was originally discussed by Carmichael in Refs.\cite{Carmichael1985,Carmichael1991}. The effect involves the quantum interference between the direct excitation of the 2-photon state of the cavity $\left| {00} \right\rangle  \to \left| {10} \right\rangle  \to \left| {20} \right\rangle$ via the pump and, the path involving the coupling to the emitter $\left| {00} \right\rangle  \to \left| {01} \right\rangle  \to \left| {11} \right\rangle  \to \left| {20} \right\rangle$. The corresponding energy level diagram and the links between the states is given in Fig.\ref{Fig:g2opt_JC}(b). Once again the antibunching condition is very restrictive when only the cavity mode is driven. Our treatment shows that the mechanism can be extended to the case where $F_2\neq0$ to allow for arbitrary system parameters. Such a configuration was implemented in Ref.\cite{Hamsen2017}, for example, where the both the cavity and the single atom are driven. In that case, we obtain the following optimal condition on the field driving the emitter
\begin{equation}
    {F_{2|{\rm{opt}}}} = {F_1}\frac{{{{\tilde \Delta }_1} + {{\tilde \Delta }_2} \pm \sqrt {{{\tilde \Delta }_1}\left( {{{\tilde \Delta }_1} + {{\tilde \Delta }_2}} \right) - {g^2}} }}{g}\,.
\end{equation}
We show in Fig.\ref{Fig:g2opt_JC}(c) an example of second order correlation function of the cavity field as a function of its occupation, under the conditions of Eqs.(\ref{D1opt_JC},\ref{gopt_JC}). We recover the linear increase obtained in the case of the coupled Kerr cavities. The sub-Poissonian statistics however breaks down at lower occupation in that case. In panel (b), we have computed the corresponding two-time correlation $g^{(2)}_c(\tau)$ -- for a mean occupation of $n_c\simeq10^{-2}$ [red square in panel (c)] -- which displays smooth variations given that $g<\kappa$.

\section{Conclusions}\label{Sec:Conclusions}
The UPB could be suitably implemented in most nonlinear quantum photonics systems, where coupled modes and/or polarization degeneracies are available, provided that dephasing is sufficiently small.

Among the most promising systems, we have mentioned optimized Silicon photonic crystal cavities \cite{Dharanipathy2014} which present very low footprints, operate at room temperature and are highly integrable. Moreover, the UPB does not require a quantum dot and  was shown to require very low input power as opposed to heralded sources for similar repetition rates of a few MHz \cite{Flayac2015}. While coupled cavities are easily engineered \cite{Chalcraft2011,Deotare2009,Sato2012}, some simpler configurations involving a single cavity could be envisaged. Indeed since, only one of the two mode must host a finite nonlinearity \cite{Bamba2011,Flayac2016} one could implement a scheme where the cavity field is mixed with that of a properly designed waveguide \cite{Xu2014a,Fischer2017}.

Superconducting quantum circuits \cite{Gu2017} are seen nowadays as the most serious contender for quantum computation schemes. In such systems, the effective Kerr nonlinearity or coupling between the circuit and microwave photons can be tuned in wide ranges while, at the same time, the signal to noise ratio is extremely small. A suitable configuration towards a proof-of-principle of the UPB could be that of Ref.\cite{Eichler2014} where coupled nonlinear modes where engineered. However as mentioned in Sec.\ref{Sec:CWDrive} at microwave wavelengths, the unavoidable presence of thermal photons impose constraints on the minimum intracavity occupation.

As we have shown in Sec.\ref{Sec:Alternative}, the UPB can also be used to obtain a sub-Poissonian statistics in a weakly coupled cavity QED system. It would e.g. relax the requirement for a finely positioned two-level emitter inside the cavity. Besides, the UPB can even be considered to enhance antibunching in the strong coupling regime \cite{Tang2015,Radulaski2017}.

Finally, we have mentioned the potential of the UPB to uncover nonclassical signatures in semiconductor microcavities \cite{Laussy2017} assisted by the excitonic interactions. One could either rely on spatially coupled polariton modes \cite{Abbarchi2013,Adiyatullin2017} or on the polarization degree of freedom \cite{Shelykh2010a}. In the latter case the input-output mixing, we described in Sec.\ref{Sec:IOTheory}, would be naturally realized by varying the driving and detection polarization. As a result a single micropillar would be sufficient to achieve the UPB. In semiconductor microcavities, the typical photon losses occur on a picosecond timescale which paves the way to high emission rates in the GHz to THz range despite a low intracavity occupation.

The UPB concept can be extended to several other schemes. In particular, it was shown to occur in parametrically coupled modes \cite{Liew2010} or three coupled cQED systems \cite{Bamba2011} and even to induce entanglement \cite{Liew2013} in the presence of a weak nonlinearity. In general, quantum interferences can be engineered to occur for arbitrary photon numbers. Noteworthy is the suppression of the one photon probability inducing a strong bunching and favoring photon pairs which could turn out to be beneficial for potential heralded schemes. One could even think of networks for which the probability distribution is fully tailored. Finally the UPB could be exploited for conventional single photon protocols, e.g. to optimize the performances of single photon sources based on four-wave mixing or parametric down conversion.

In conclusion, we have reviewed the unconventional photon blockade mechanism that can be interpreted in terms of quantum interference or optimal squeezing. We have shown how a proper mixing of the input and output fields allows a measurable antibunching to be obtained for arbitrary system parameters. In particular the output mixing allows consideration of fully decoupled nonlinear modes which could turn into a great advantage for the experimental realization of the effect. In particular it allows to consider for instance a single cavity mode with polarization degeneracy. Finally, we have discussed alternative systems where the unconventional blockade can be transposed and the ensuing applications.

\bibliography{Bibliography}

\end{document}